\documentclass[a4paper,10pt,notitlepage]{article}

\usepackage[latin1]{inputenc}
\usepackage{amsmath}
\usepackage{amssymb}
\usepackage{bm}
\usepackage{graphicx}
\usepackage{lscape}
\usepackage{setspace}
\doublespace
\usepackage{lineno}
\linenumbers

\usepackage{jpc}

\def\0{\phantom{0}}
\def\.{\phantom{.}}
\def\my0{\phantom{.}0\phantom{.0000}}

\title{A set of molecular models based on quantum mechanical \emph{ab initio} calculations and thermodynamic data}
\author{Bernhard Eckl, Jadran Vrabec\footnote{Tel.: +49-711/685-66107, Fax: +49-711/685-66140, Email: vrabec@itt.uni-stuttgart.de}, and Hans Hasse}
\date{Institut f\"ur Technische Thermodynamik und Thermische Verfahrenstechnik, Universit\"at Stuttgart \\
Pfaffenwaldring 9, 70550 Stuttgart, Germany}

\begin{document}
\maketitle

\begin{abstract}
A parameterization strategy for molecular models on the basis of force fields is proposed, which allows a rapid development of models for small molecules by using results from quantum mechanical (QM) \emph{ab initio} calculations and thermodynamic data. The geometry of the molecular models is specified according to the atom positions determined by QM energy minimization. The electrostatic interactions are modeled by reducing the electron density distribution to point dipoles and point quadrupoles located in the center of mass of the molecules. Dispersive and repulsive interactions are described by Lennard-Jones sites, for which the parameters are iteratively optimized to experimental vapor-liquid equilibrium (VLE) data, i.e. vapor pressure, saturated liquid density, and enthalpy of vaporization of the considered substance. The proposed modeling strategy was applied to a sample set of ten molecules from different substance classes. New molecular models are presented for iso-butane, cyclohexane, formaldehyde, dimethyl ether, sulfur dioxide, dimethyl sulfide, thiophene, hydrogen cyanide, acetonitrile, and nitromethane. Most of the models are able to describe the experimental VLE data with deviations of a few percent.
\end{abstract}

\vspace{5mm}
\noindent {\bfseries Keywords:} Molecular modeling; modeling strategy; vapor-liquid equilibrium; critical properties; iso-butane; cyclohexane; formaldehyde; dimethyl ether; sulfur dioxide; dimethyl sulfide; thiophene; hydrogen cyanide; acetonitrile; nitromethane

\section{Introduction}
\label{Einleitung}
Molecular modeling and simulation is a progressive approach for describing and predicting thermophysical properties of both pure substances and mixtures of technical interest. Many authors have shown the excellent capabilities of this molecular approach for different applications \cite{Ungerer2007, Ungerer2007a}.  Unfortunately, the more widespread use of molecular methods for engineering applications is still restricted by the scarcity of suitable molecular models.

For many substances, transferable molecular models have been developed, i.e. force fields for classes of compounds like alkanes or alcohols. Thereby, it is assumed that the parameters for functional groups are valid for different molecular species. The main disadvantage of transferable models is that their parameters were adjusted for the whole substance class and may not be optimal for a specific substance. Furthermore, they do not cover substances outside the modeled class. An overview and assessment of some commonly used transferable potentials can be found in \cite{Martin2006}.

Molecular models that were optimized for specific substances are available only for selected compounds and, particularly older ones, sometimes do not show the desired quality. Therefore, from an engineering point of view, a method for the rapid development of new molecular models for specific substances is of great interest. Moreover, the molecular models should be accurate, simple, and computationally efficient. In the present paper, a systematic strategy is proposed to develop such models.

The present work is restricted to rigid, non-polarizable models for comparatively small molecules. The models have state-independent parameters throughout. For computational efficiency, the united-atom approach is used, i.e. hydrogen atoms bonded to carbon are not modeled explicitly. The proposed strategy uses information determined by quantum mechanical (QM) \emph{ab initio} calculations to include physically sound molecular properties and to reduce the number of adjustable parameters. A remaining subset of model parameters -- typically two to four -- is subsequently optimized by adjustment to experimental data on vapor-liquid equilibria (VLE) of the pure substances. The aim is to achieve deviations to experimental values for the vapor pressure, saturated liquid density, and enthalpy of vaporization in the range from triple point to critical point of below 5, 1, and 5\%, respectively.

Such accurate models are known to have an excellent extrapolative and predictive power. This was recently shown, e.g., by the prediction of 17 different thermophysical properties for ethylene oxide, covering phase equilibria, thermal, caloric, transport properties, and surface tension \cite{Eckl2007}, or for ammonia including structural quantities \cite{Eckl2008}.

Molecular models that were developed on the basis of QM calculations stand between strictly \emph{ab initio} models and fully empirical models. The present strategy is based on the idea to include substantial \emph{ab initio} information without giving up the freedom to reasonably optimize the model to important macroscopic thermodynamic properties. Thus, for the modeling process some experimental data is needed for optimization. All three chosen properties, mentioned above, have the advantage to be well available for numerous engineering fluids and to represent dominant features of the fluid state.

The parameters of a molecular model can be separated into three groups. Firstly, geometric parameters specify the locations of the different interaction sites of the molecular model. Secondly, electrostatic parameters define the interactions of static polarities of the single molecules. And finally, dispersive and repulsive parameters determine the attraction by London forces and the repulsion by overlaps of the electronic orbitals. Here, the Lennard-Jones 12-6 (LJ) potential \cite{Jones1924, Jones1924a} was used to assure straightforward compatibility with the overwhelming majority of the molecular models in the literature.

To describe the intermolecular interactions, a varying number of LJ sites and superimposed ideal point dipoles and/or ideal linear point quadrupoles were used. Point dipoles or quadrupoles were employed for the description of the electrostatic interactions to reduce the computational effort significantly. A point dipole may, e.g. when a simulation program does not support this interaction site type, be approximated by two point charges $\pm q$ separated by a distance $l$. Limited to small $l$, one is free to choose this distance as long as $\mu = q l$ holds. Analogously, a point quadrupole can be approximated by three collinear point charges $q$, $-2q$, and $q$ separated by $l$ each, where $Q = 2 q l^2$.

The total intermolecular interaction energy writes as

\noindent
\begin{multline}\label{eq_energy}
 U = \sum_{i=1}^{N-1} \sum_{j=i+1}^N \left\lbrace \sum_{a=1}^{S_i^\mathrm{LJ}} \sum_{b=1}^{S_j^\mathrm{LJ}} 4 \varepsilon_{ijab} \left[ \left( \frac{\sigma_{ijab}}{r_{ijab}} \right)^{12} - \left( \frac{\sigma_{ijab}}{r_{ijab}} \right)^6 \right] \right. + \\
 \left. \sum_{c=1}^{S_i^\mathrm{e}} \sum_{d=1}^{S_j^\mathrm{e}} \frac{1}{4 \pi \epsilon_0} \left[ \frac{\mu_{ic} \mu_{jd}}{r_{ijcd}^3} \cdot f_1(\bm{\omega}_i, \bm{\omega}_j) + \frac{\mu_{ic} Q_{jd} + Q_{ic} \mu_{jd}}{r_{ijcd}^4} \cdot f_2(\bm{\omega}_i, \bm{\omega}_j) + \frac{Q_{ic} Q_{jd}}{r_{ijcd}^5} \cdot f_3(\bm{\omega}_i, \bm{\omega}_j) \right] \right\rbrace,
\end{multline}

\noindent where $r_{ijab}$, $\varepsilon_{ijab}$, $\sigma_{ijab}$ are the distance, the LJ energy parameter, and the LJ size parameter, respectively, for the pair-wise interaction between LJ site $a$ on molecule $i$ and LJ site $b$ on molecule $j$. $\epsilon_0$ is the permittivity of vacuum, whereas $\mu_{ic}$ and $Q_{ic}$ denote the dipole moment and the quadrupole moment of the electrostatic interaction site $c$ on molecule $i$, and so forth. $f_x(\bm{\omega}_i, \bm{\omega}_j)$ are expressions for the dependency of the electrostatic interactions on the orientations $\bm{\omega}_i$ and $\bm{\omega}_j$ of the molecules $i$ and $j$, cf. \cite{Allen1987, Gray1984}. Finally, the summation limits $N$, $S_x^\mathrm{LJ}$, and $S_x^\mathrm{e}$ denote the number of molecules, the number of LJ sites, and the number of electrostatic sites, respectively.

Interactions between LJ sites of different type are determined by applying the standard Lorentz-Berthelot combining rules \cite{Lorentz1881, Berthelot1898}

\noindent
\begin{equation}
 \sigma_{ijab} = \frac{\sigma_{iiaa} + \sigma_{jjbb}}{2},
\end{equation}

\noindent and

\noindent
\begin{equation}
 \varepsilon_{ijab} = \sqrt{\varepsilon_{iiaa} \varepsilon_{jjbb}}.
\end{equation}

\section{Molecular properties from QM}
\label{QM}
In a recent publication, Sandler et al. \cite{Sandler2007} gives a brief overview on the use of QM for the calculation of thermophysical properties. By numerically solving Schr\"odinger's equation, it is nowadays possible to calculate different molecular properties for technically relevant components in a quite standardized way. Many different QM codes are available for this task. For license reasons, the open source code GAMESS(US) \cite{Schmidt1993} was used in the present work.

\subsection{Geometry}
All geometric data of the molecular models, i.e. bond lengths, angles, and dihedrals, were directly taken from QM calculations. Therefore, a geometry optimization, i.e. an energy minimization, was initially performed using GAMESS(US) \cite{Schmidt1993}. The Hartree-Fock level of theory was applied with a relatively small (6-31G) basis set. Alternatively, density functional theory (DFT) methods, e.g. BLY3P, can be used, as they are known to give reasonable results for the molecular structure \cite{Leach2001}.

\begin{figure}[ht]
\includegraphics[scale=0.5]{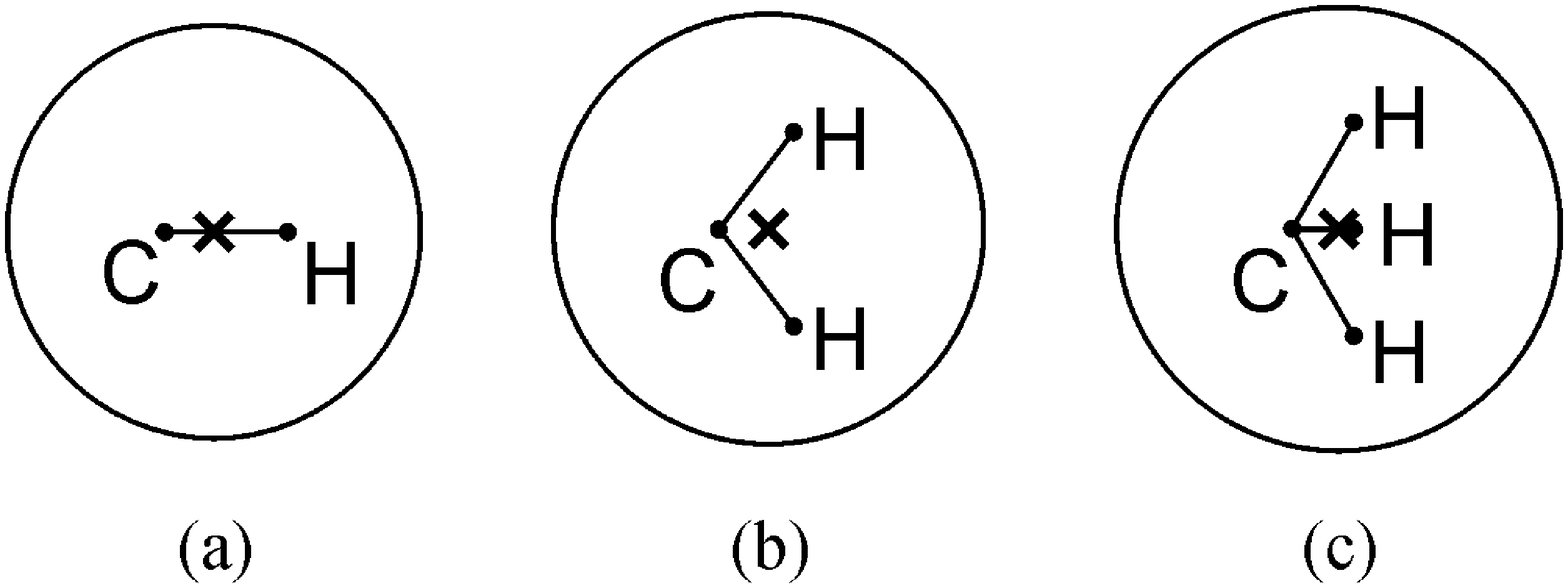}
\bigskip
\caption[Geometry of the united-atom sites: (a)~The methine (CH) site is located at 0.4 of the carbon-hydrogen distance, (b)~the methylene (CH$_2$) site is located at the geometric mean, and (c)~the methyl (CH$_3$) site is located at the geometric mean.]{\label{fig_united_atom}Eckl et al.} 
\end{figure}

The resulting configuration of the atoms was taken without subsequent modification to specify the position of the LJ sites in space, except for the hydrogen atoms. As the united atom approach was used to obtain computationally efficient molecular models, the hydrogen atoms were modeled together with the carbon atom they are bonded to. For the methylene (CH$_2$) and methyl (CH$_3$) united atom sites, the LJ potential was located at the geometric mean of the nuclei, while the methine (CH) united atom site was located at 0.4 of the distance between carbon and hydrogen atom, cf. Figure~\ref{fig_united_atom}. These empirical offsets are in good agreement with the results of Ungerer et al. \cite{Ungerer2000}, which were found by optimization of transferable molecular models for n-alkanes.

\subsection{Electrostatics}
Intermolecular electrostatic interactions mainly occur due to static polarities of single molecules that can well be obtained by QM. Here, the M\o{}ller-Plesset 2 level was used that considers electron correlation in combination with the polarizable 6-311G(d,p) basis set.

The purpose of the present work is the development of effective pair potentials with a state-independent set of model parameters. Obviously, the electrostatic interactions are stronger in the liquid state than in the gaseous state due to the higher density. Furthermore, the mutual polarization raises their magnitude in the liquid. Thus, for the calculation of the electrostatic moments by QM a liquid-like state should be considered. This was done here by placing the molecule within a dielectric continuum and assigning the experimental dielectric constant of the liquid to it, as in the COSMO method \cite{Klamt1995}.

From the resulting electron density distribution for the small symmetric molecules regarded here, ideal point dipoles and ideal linear point quadrupoles were estimated by simple integration over the orbitals. Magnitudes and orientations of these electrostatic interaction sites were used in the molecular models without any modification.

For other, more complex molecules, more sophisticated methods like CHELP \cite{Chirlian1987}, CHELPG \cite{Breneman1990}, or the distributed multipole analysis \cite{Stone2002} are available in the literature. These methods adjust a set of partial charges or higher order electrostatic sites to the electrostatic potential around the molecule calculated by QM. Although they are able to reflect the electrostatic interactions with higher accuracy, they are not considered here. They always yield a larger number of interaction sites if they are not co-located with other sites and would thus lead to computationally more expensive molecular models.

\subsection{Dispersion and Repulsion}
It would be highly desirable to also calculate the dispersive and repulsive interactions using \emph{ab initio} methods as well. This approach was followed by different authors in the past, e.g. for neon \cite{Eggenberger1994, Vogt2001, Garrison2002, Nasrabad2004}, argon \cite{Vogt2001, Nasrabad2004, Ermakova1995}, krypton \cite{Nasrabad2003}, nitrogen \cite{Leonhard2002}, carbon dioxide \cite{Welker1996}, hydrogen chloride \cite{Naicker2003}, acetonitrile \cite{Hloucha2000}, methanol \cite{Hloucha2000}, acetylene \cite{Garrison2004}, and methanethiol \cite{Garrison2005}. However, from an engineering point of view, this leads to difficulties.

For an estimation of dispersive and repulsive interactions at least two molecules must be taken into account. To properly scan the energy hyper surface, many QM calculations at different distances and orientations of the molecules have to be performed. As the dispersive, and partly also the repulsive, interactions are just a very small fraction of the total energy calculated by QM, highly accurate methods like coupled cluster (CC) with large basis sets or even extrapolations to the basis set limit must be used for this task \cite{Sandler2007}.

Due to the fact that this is computationally too expensive for engineering purposes, we used the parameters for the dispersive and repulsive interactions for an initial model from similar sites of other molecular models. Some of these parameters were subsequently fitted in the optimization process to yield the correct VLE behavior of the modeled substance.

\section{Optimization to VLE data}
\label{Optimierung}
The optimization was performed using a Newton scheme as proposed by Stoll \cite{Stoll2005}. The applied method has many similarities with the one published by Ungerer et al. \cite{Ungerer1999} and later on modified by Bourasseau et al. \cite{Bourasseau2003}. It relies on a least-square minimization of a weighted fitness function $\mathcal{F}$ that quantifies the deviations of simulation results from a given molecular model compared to experimental data. The weighted fitness function writes as

\noindent
\begin{equation}\label{eq_fitness}
  \mathcal{F} = \frac{1}{d} \sum_{i=1}^d \frac{1}{(\delta A_{i,\mathrm{sim}})^2} (A_{i,\mathrm{sim}}(\bm{M}_0) - A_{i,\mathrm{exp}})^2~,
\end{equation}

\noindent wherein the $n$-dimensional vector $\bm{M}_0 = (m_{0,1}, ..., m_{0,n})$ represents the set of $n$ model parameters $m_{0,1}, ..., m_{0,n}$ to be optimized. The deviations of results from simulation $A_{i,\mathrm{sim}}$ to experimental data $A_{i,\mathrm{exp}}$ are weighted with the expected simulation uncertainties $\delta A_{i,\mathrm{sim}}$. Equation~(\ref{eq_fitness}) allows simultaneous adjustment of the model parameters to different thermophysical properties $A_i$ (saturated liquid densities $\rho'$, vapor pressures $p_\sigma$, and enthalpies of vaporization $\Delta h_\mathrm{v}$ at various temperatures in the present work).

The unknown functional dependence of the property $A_i$ on the model parameters is approximated by a first order Taylor series developed in the vicinity of the initial parameter set $\bm{M}_0$

\noindent
\begin{equation}
  A_{i,\mathrm{sim}}(\bm{M}_\mathrm{new}) = A_{i,\mathrm{sim}}(\bm{M}_0) + \sum_{j=1}^n \frac{\partial A_{i,\mathrm{sim}}}{\partial m_j} \cdot (m_{\mathrm{new},j} - m_{0,j})~.
\end{equation}

\noindent Therein, the partial derivatives of $A_i$ with respect to each model parameter $m_j$, i.e. the sensitivities, are calculated from difference quotients

\noindent
\begin{equation}
  \frac{\partial A_{i,\mathrm{sim}}}{\partial m_j} \approx \frac{A_{i,\mathrm{sim}}(m_{0,1}, ..., m_{0,j}+\Delta m_j, ..., m_{0,n}) - A_{i,\mathrm{sim}}(m_{0,1}, ..., m_{0,j}, ..., m_{0,n})}{\Delta m_j}~.
\end{equation}

Assuming a sound choice of the model parameter variations $\Delta m_j$, i.e. small enough to ensure linearity and large enough to yield differences in the simulation results significantly above the statistical uncertainties, this method allows a step-wise optimization of the molecular model by minimization of the fitness function $\mathcal{F}$. Experience shows that an optimized set of model parameters was usually found within a few iterative steps when starting from a reasonable initial model.

Correlations for vapor pressure, saturated liquid density, and enthalpy of vaporization, taken from the DIPPR database \cite{DIPPR2006}, were used as "experimental data" for model adjustment and evaluation. This was done even in cases where the correlation is based on no or only few true experimental data points, as the correlations were regarded best practice. The comparison between simulation results and experiment was done by applying fits to the simulation data according to Lotfi et al. \cite{Lotfi1992}. The relative deviation between fit and correlation was calculated in steps of 1~K from 55 to 97\% of the critical temperature and is denoted by "mean unsigned error" in the following.

Vapor-liquid equilibrium simulations were performed using the Grand Equilibrium method by Vrabec et al. \cite{Vrabec2002}, technical simulation details are given in the appendix.

\section{Molecular Models}
The selected ten molecules belong to different substance classes to show the wide applicability of the proposed strategy. In the present work, we restricted ourselves to small molecules, where the internal degrees of freedom may be neglected. Thus, the molecular models are rigid, using the most stable configuration determined by QM.

The optimized parameter sets of the new molecular models are summarized in Table~\ref{tab_Parameter}. Table~\ref{tab_critical} compares the critical properties from simulation to experimental data. The critical properties from simulation were obtained through fits to VLE simulation results as suggested by Lotfi et al. \cite{Lotfi1992}. The estimated uncertainties of critical temperature, critical density, and critical pressure from simulation are 0.5, 2, and 2\%, respectively. A very good agreement between simulation and experiment was reached, being predominantly within the combined error bars.

In the following sections, substance specific details are briefly discussed. Furthermore,
references to alternative models from the literature are given and the simulation results from the
present models are compared to simulation data from the literature where available. Numerical VLE
simulation results are given as Supplemtary Material.

\subsection{Iso-Butane and Cyclohexane}
The branched alkane iso-butane and the cyclic alkane cyclohexane show only very weak static polarities. Here, the main contributions to the intermolecular interaction are dispersion and repulsion. The electrostatic interactions have only a minor influence but should not be neglected completely.

In the literature different molecular models for iso-butane can be found, which are mostly based on force fields for branched alkanes. The well-known OPLS force field by Jorgensen et al. is available in two versions for iso-butane, one using the united-atom approach \cite{Jorgensen1984} and one using an all-atom description \cite{Jorgensen1996}. Both OPLS force fields were optimized to liquid density and enthalpy of vaporization at 293~K only. The model of Poncela et al. \cite{Poncela1997} was adjusted to yield correct second virial coefficients. For a better applicability in a wider range of states, recent developments were optimized to experimental VLE data. Examples from this group are the force fields presented by Nath and de Pablo \cite{Nath2000}, Martin and Siepmann \cite{Martin1999}, Bourasseau et al. \cite{Bourasseau2002}, or Chang and Sandler \cite{Chang2004}.

For the present model, four LJ sites, one for each methyl group and one for the methine group, were used to describe dispersion and repulsion of iso-butane. The polarity was modeled by a single (weak) dipole (0.1347~D) located in the center of mass. Orientation and magnitude of the dipole were taken from the QM calculation. For the initial model, the LJ parameters were taken from Ungerer et al. \cite{Ungerer2000}. It was sufficient to adjust a single parameter, the offset distance of the methine group. It was optimized to 0.4 of the carbon-hydrogen distance, cf. Figure~\ref{fig_united_atom} and held constant subsequently. The parameters of the present model are given in Table~\ref{tab_Parameter}.

\begin{figure}[ht]
\includegraphics[scale=0.5]{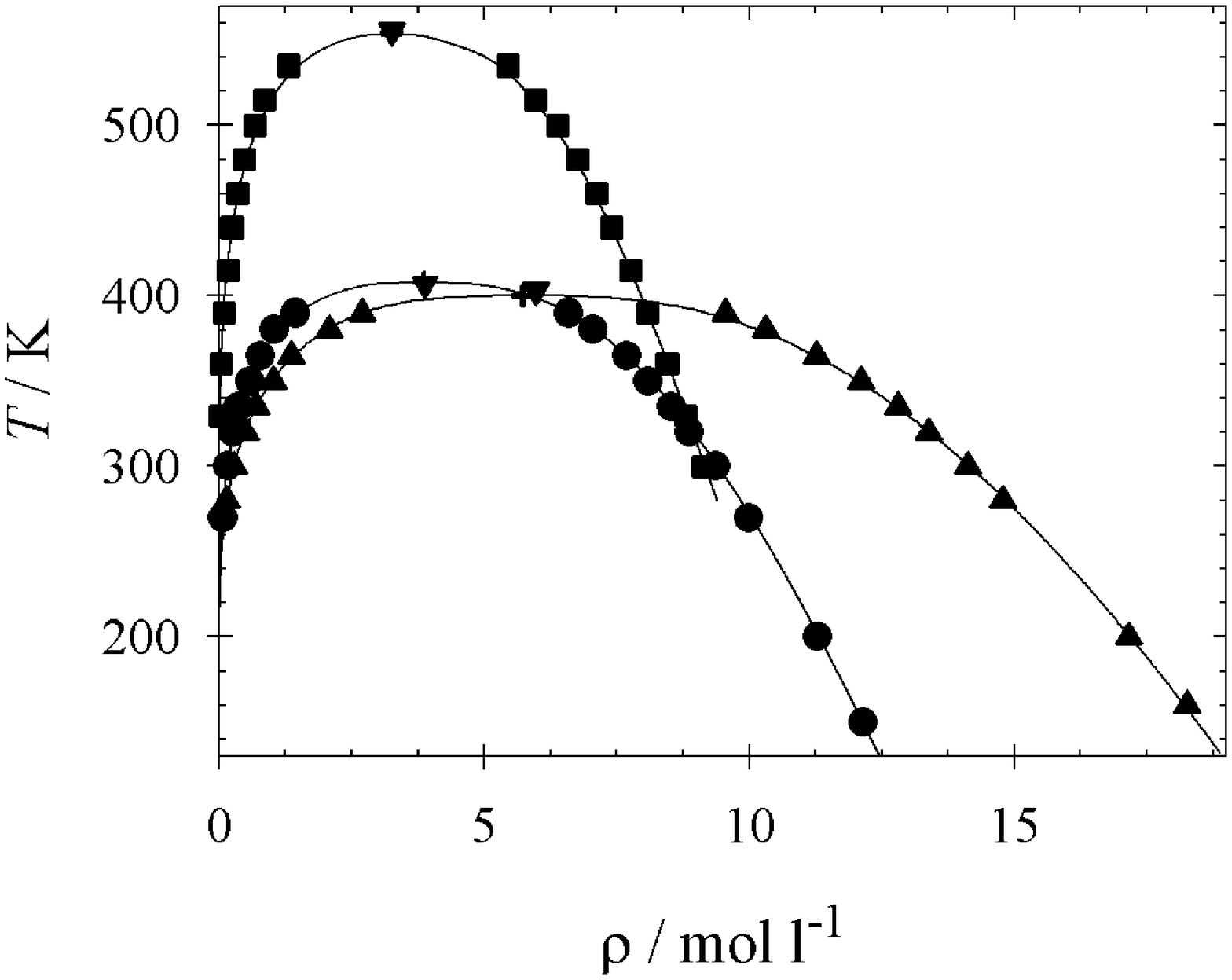}
\bigskip
\caption[Saturated densities. Present simulation data: {\Large $\bullet$}~iso-butane, {\small $\square$}~cyclohexane, {\Large $\circ$}~dimethyl ether. {---}~correlations of experimental data \cite{DIPPR2006}, +~critical points derived from simulation, $\blacktriangledown$~experimental critical points.]{\label{fig_vle1_rho}Eckl et al.} 
\end{figure}

\begin{figure}[ht]
\includegraphics[scale=0.5]{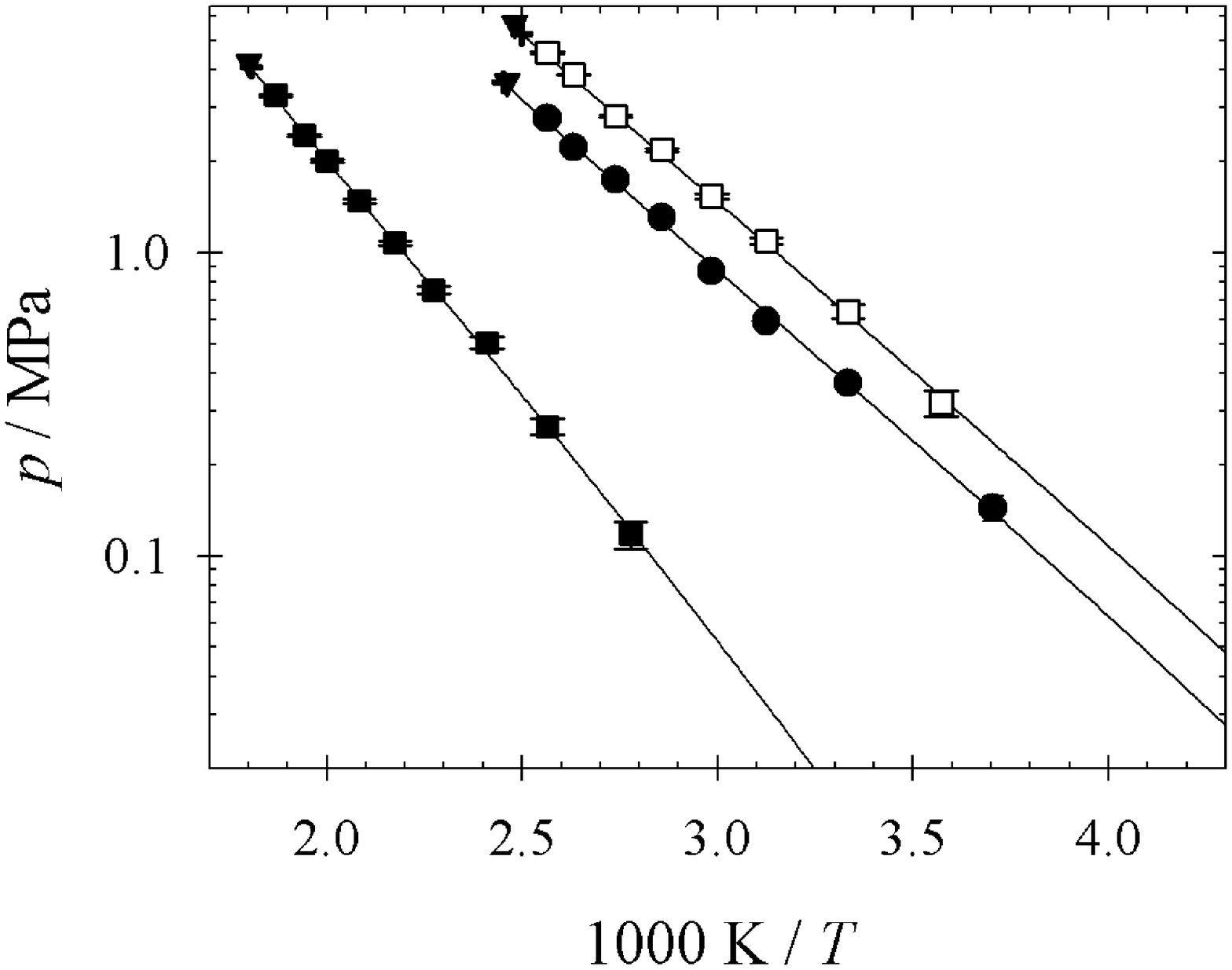}
\bigskip
\caption[Vapor pressure. Present simulation data: {\Large $\bullet$}~iso-butane, {\small $\blacksquare$}~cyclohexane, {\Large $\circ$}~dimethyl ether. {---}~correlations of experimental data \cite{DIPPR2006}, +~critical points derived from simulation, $\blacktriangledown$~experimental critical points.]{\label{fig_vle1_p}Eckl et al.}
\end{figure}

\begin{figure}[ht]
\includegraphics[scale=0.5]{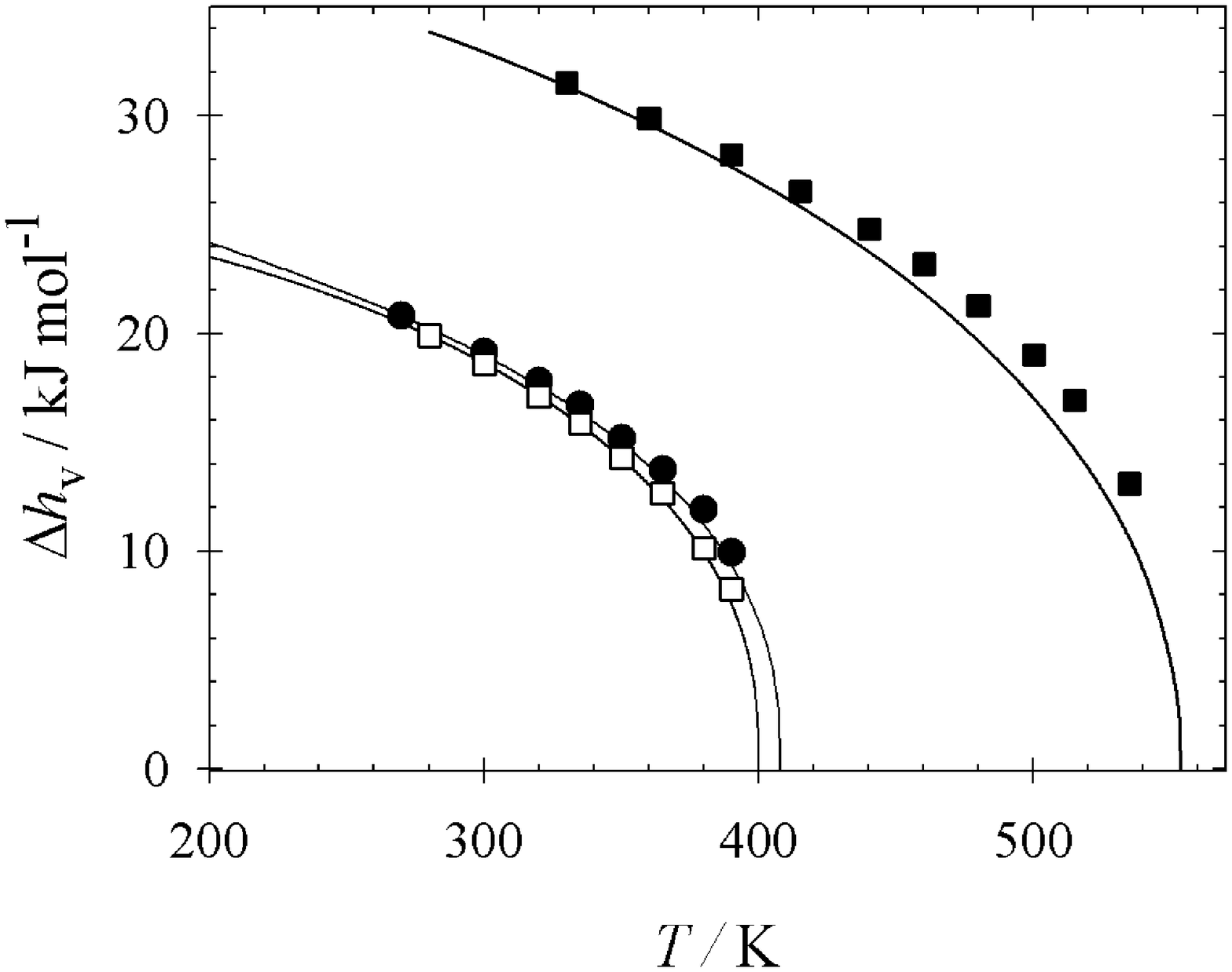}
\bigskip
\caption[Enthalpy of vaporization. Present simulation data: {\Large $\bullet$}~iso-butane, {\small $\blacksquare$}~cyclohexane, {\Large $\circ$}~dimethyl ether. {---}~correlations of experimental data \cite{DIPPR2006}.]{\label{fig_vle1_dhv}Eckl et al.}
\end{figure}

\begin{figure}[ht]
\includegraphics[scale=0.5]{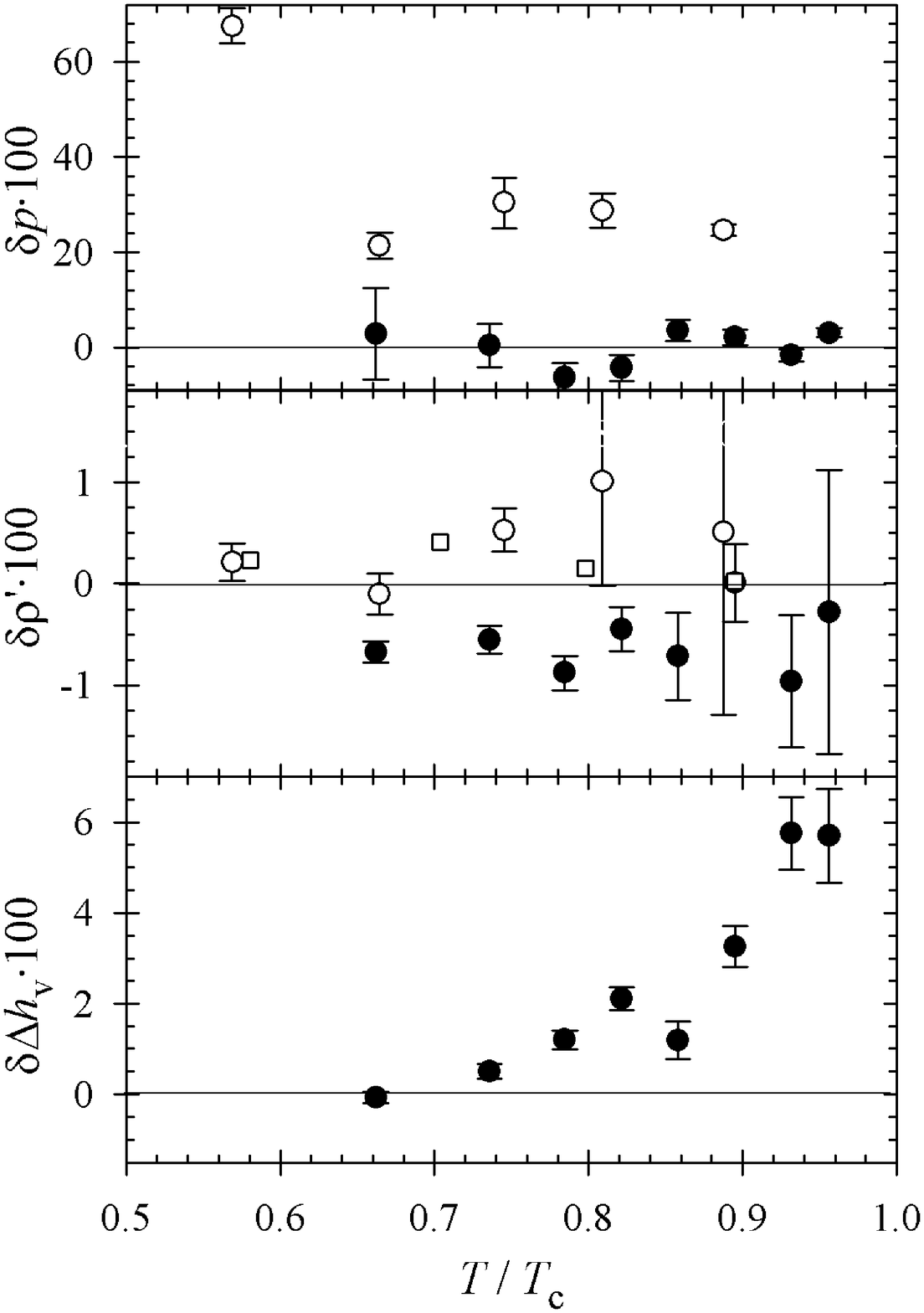}
\bigskip
\caption[Relative deviations of vapor-liquid equilibrium properties between simulation data and correlations of experimental data \cite{DIPPR2006} ($\delta z=(z_\mathrm{sim}-z_\mathrm{exp})/z_\mathrm{exp}$) for iso-butane: {\Large $\bullet$}~present model, {\Large $\circ$}~Martin and Siepmann \cite{Martin1999}, {\small $\square$}~Nath and de Pablo \cite{Nath2000}. Top: vapor pressure, center: saturated liquid density, bottom: enthalpy of vaporization.]{\label{fig_vle_ibutan_dev}Eckl et al.}
\end{figure}

Figures~\ref{fig_vle1_rho} to \ref{fig_vle1_dhv} show saturated densities, vapor pressure, and enthalpy of vaporization, respectively, from the present iso-butane model in comparison to experimental data \cite{DIPPR2006}. Figure~\ref{fig_vle_ibutan_dev} shows a deviation plot between simulation and experimental data. In the deviation plot also simulation results from Martin and Siepmann, using their TraPPE force field, \cite{Martin1999} and from Nath and de Pablo \cite{Nath2000} are included. A very good agreement was obtained for the present model yielding mean unsigned errors in vapor pressure, saturated liquid density, and enthalpy of vaporization of 4.2, 0.6, and 1.8\%, respectively, in the temperature range from 55 to 97\% of the critical temperature, which is about 225 to 395~K. For vapor pressure, the present model yields significantly better results than the TraPPE force field, while no simulation data is available from Nath and de Pablo for this property. For saturated liquid density, all three models yield comparable results within 1\% deviation to experimental data. No comparison between the models was possible for enthalpy of vaporization due to missing numerical data in \cite{Nath2000, Martin1999}.

For cyclohexane different molecular models are available in the literature \cite{Errington1999, Neubauer1999, Faller1999, Bourasseau2002a} which all account for the internal degrees of freedom. Nevertheless, the present cyclohexane model was assumed to be rigid and in its most stable configuration, i.e. the saddle shape. The molecular model consists of six LJ sites, one for each methylene group. The static polarity was modeled by a single quadrupole parameterized according to QM. The two LJ parameters for methylene were optimized to experimental VLE data. The parameters of the present model are given in Table~\ref{tab_Parameter}.

\begin{figure}[ht]
\includegraphics[scale=0.5]{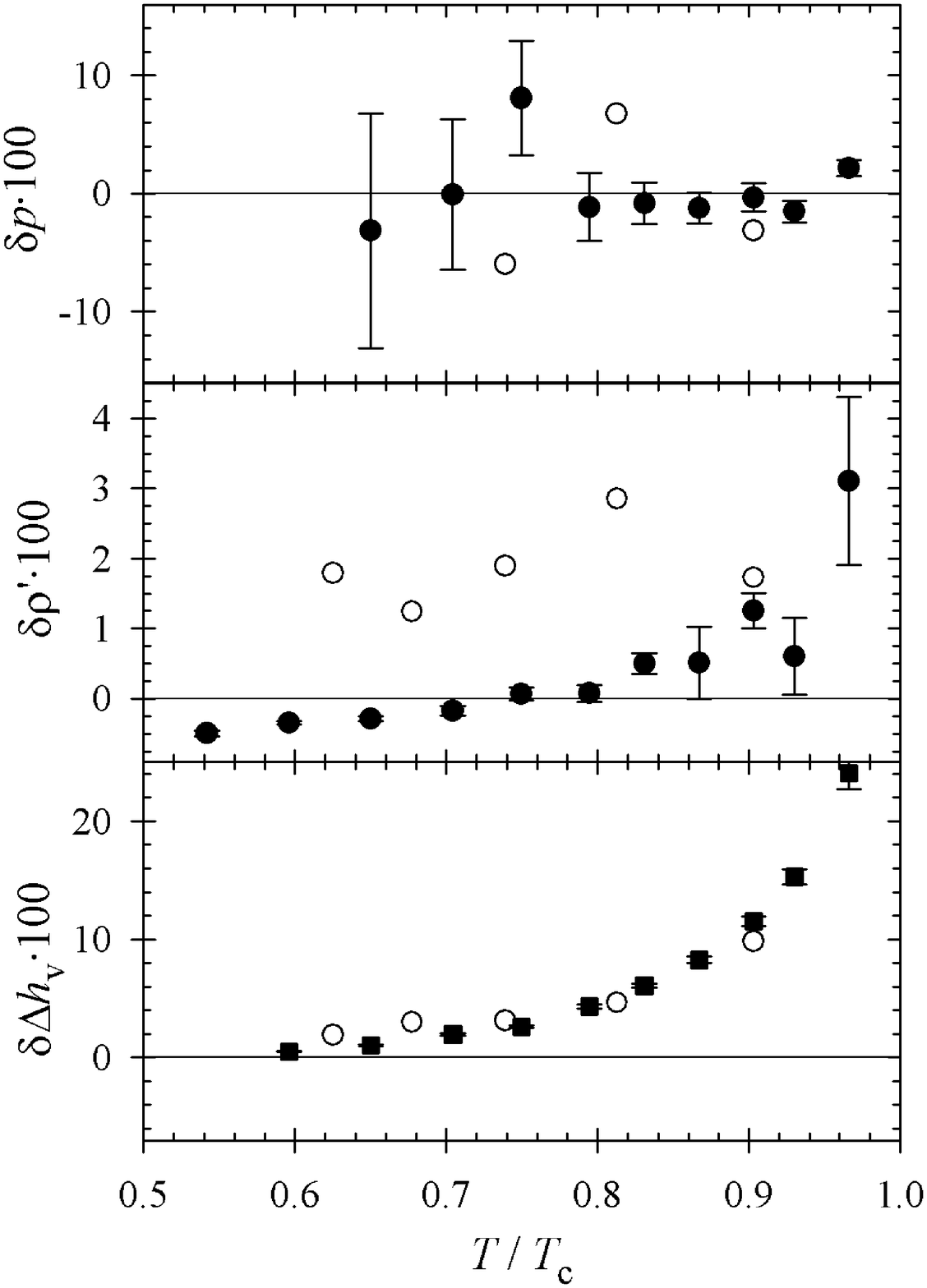}
\bigskip
\caption[Relative deviations of vapor-liquid equilibrium properties between simulation data and correlations of experimental data \cite{DIPPR2006} ($\delta z=(z_\mathrm{sim}-z_\mathrm{exp})/z_\mathrm{exp}$) for cyclohexane: {\Large $\bullet$}~present model, {\Large $\circ$}~Bourasseau et al. \cite{Bourasseau2002}. Top: vapor pressure, center: saturated liquid density, bottom: enthalpy of vaporization.]{\label{fig_vle_chex_dev}Eckl et al.}
\end{figure}

Figures~\ref{fig_vle1_rho} to \ref{fig_vle1_dhv} again show present VLE simulation results in comparison to experimental data for cyclohexane. Figure~\ref{fig_vle_chex_dev} shows the deviation plot including simulation results from Bourasseau et al. \cite{Bourasseau2002}. The mean unsigned errors in vapor pressure, saturated liquid density, and enthalpy of vaporization for the present model are 0.9, 0.5, and 5.6\%, respectively. Simulation results for the enthalpy of vaporization show systematic relative deviations towards the critical point, while the absolute deviation is below 2.5~kJ/mol.

Compared to the model of Bourasseau et al. improvements in the description of the saturated liquid density were achieved. The simulation results for vapor pressure agree within their assumed simulation uncertainties which were not reported by Bourasseau et al. An interesting point is that both molecular models yield the same deviations from experimental data on the enthalpy of vaporization, while the DIPPR database reports true experimental data up to $0.97 T_\mathrm{c}$. For all other molecular models for cyclohexane mentioned above, no numerical VLE simulation data is reported in the literature. Thus, no comparison is made here.

\subsection{Formaldehyde and Dimethyl Ether}
The present molecular model for formaldehyde consists of two LJ sites, one for the oxygen atom and one for the methylene group, as well as one dipole. The dipole is located in the center of mass and its moment was specified according to QM results, cf. Table~\ref{tab_Parameter}. All four LJ parameters were adjusted to experimental VLE data and are given in Table~\ref{tab_Parameter}.

\begin{figure}[ht]
\includegraphics[scale=0.5]{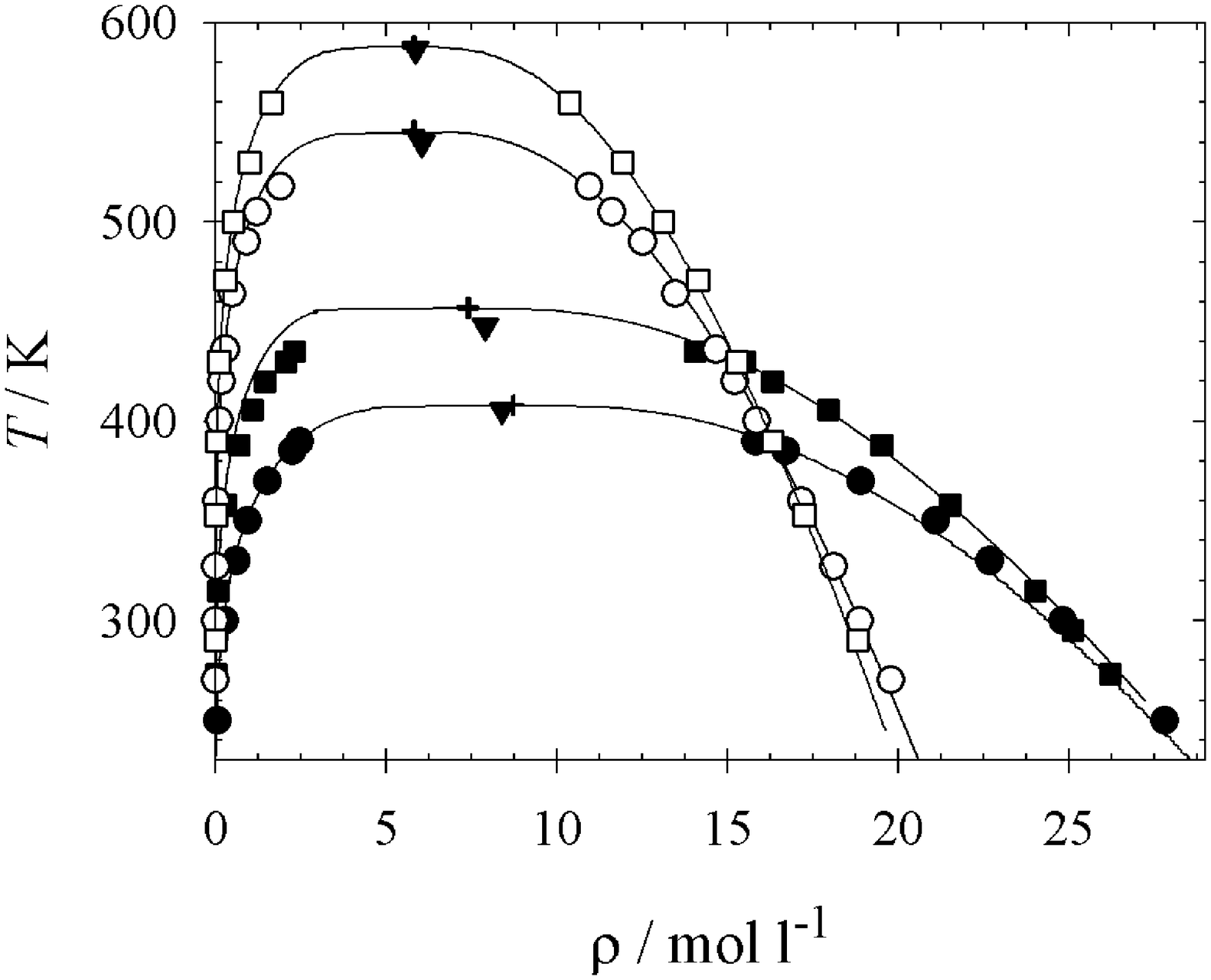}
\bigskip
\caption[Saturated densities. Present simulation data: {\Large $\bullet$}~formaldehyde, {\small $\blacksquare$}~hydrogen cyanide, $\blacklozenge$~acetonitrile, $\circ$~nitromethane. {---}~correlations of experimental data \cite{DIPPR2006}, +~critical points derived from simulation, $\blacktriangledown$~experimental critical points.]{\label{fig_vle3_rho}Eckl et al.} 
\end{figure}

\begin{figure}[ht]
\includegraphics[scale=0.5]{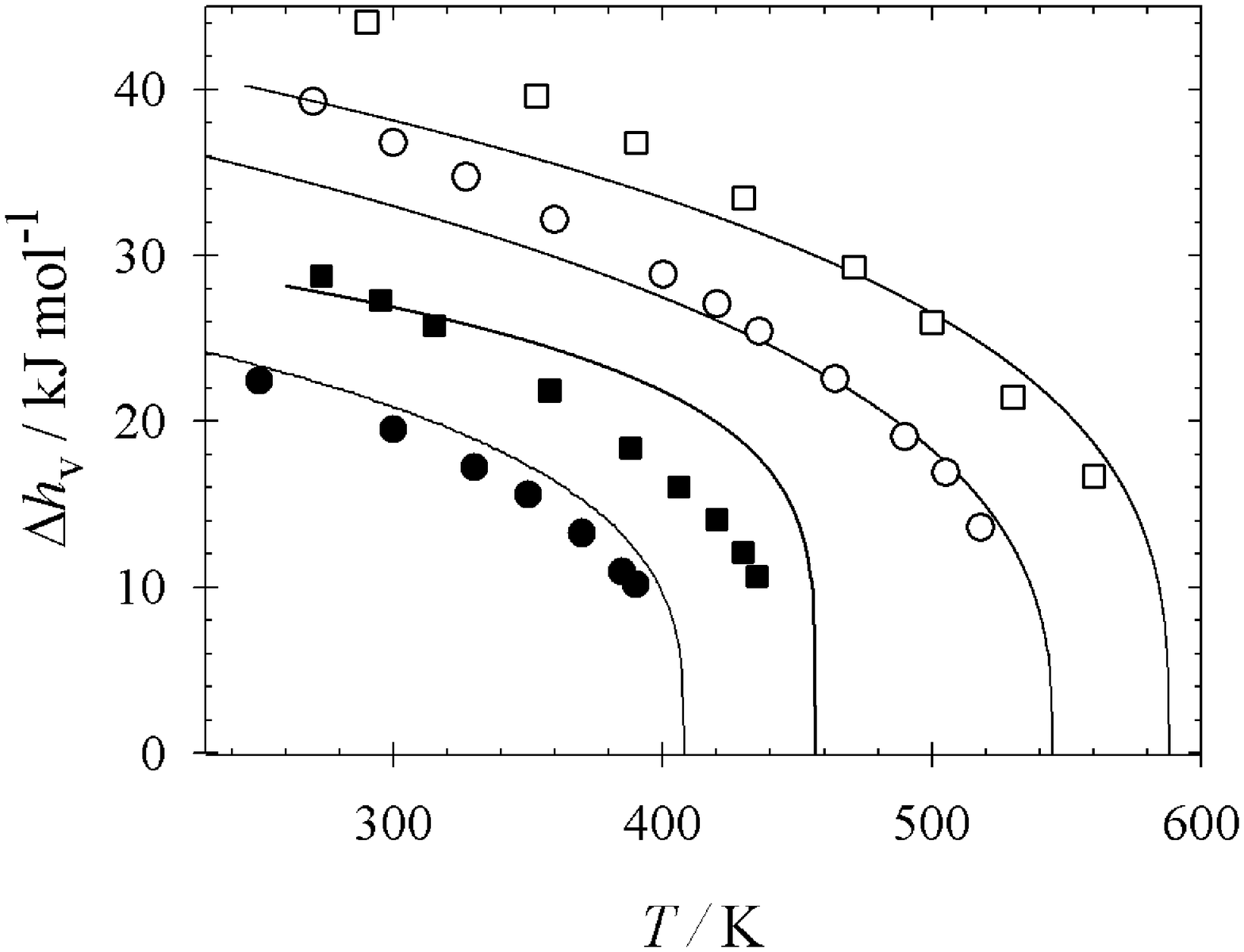}
\bigskip
\caption[Enthalpy of vaporization. Present simulation data: {\Large $\bullet$}~formaldehyde, {\small $\blacksquare$}~hydrogen cyanide, $\blacklozenge$~acetonitrile, $\circ$~nitromethane. {---}~correlations of experimental data \cite{DIPPR2006}.]{\label{fig_vle3_dhv}Eckl et al.}
\end{figure}

\begin{figure}[ht]
\includegraphics[scale=0.5]{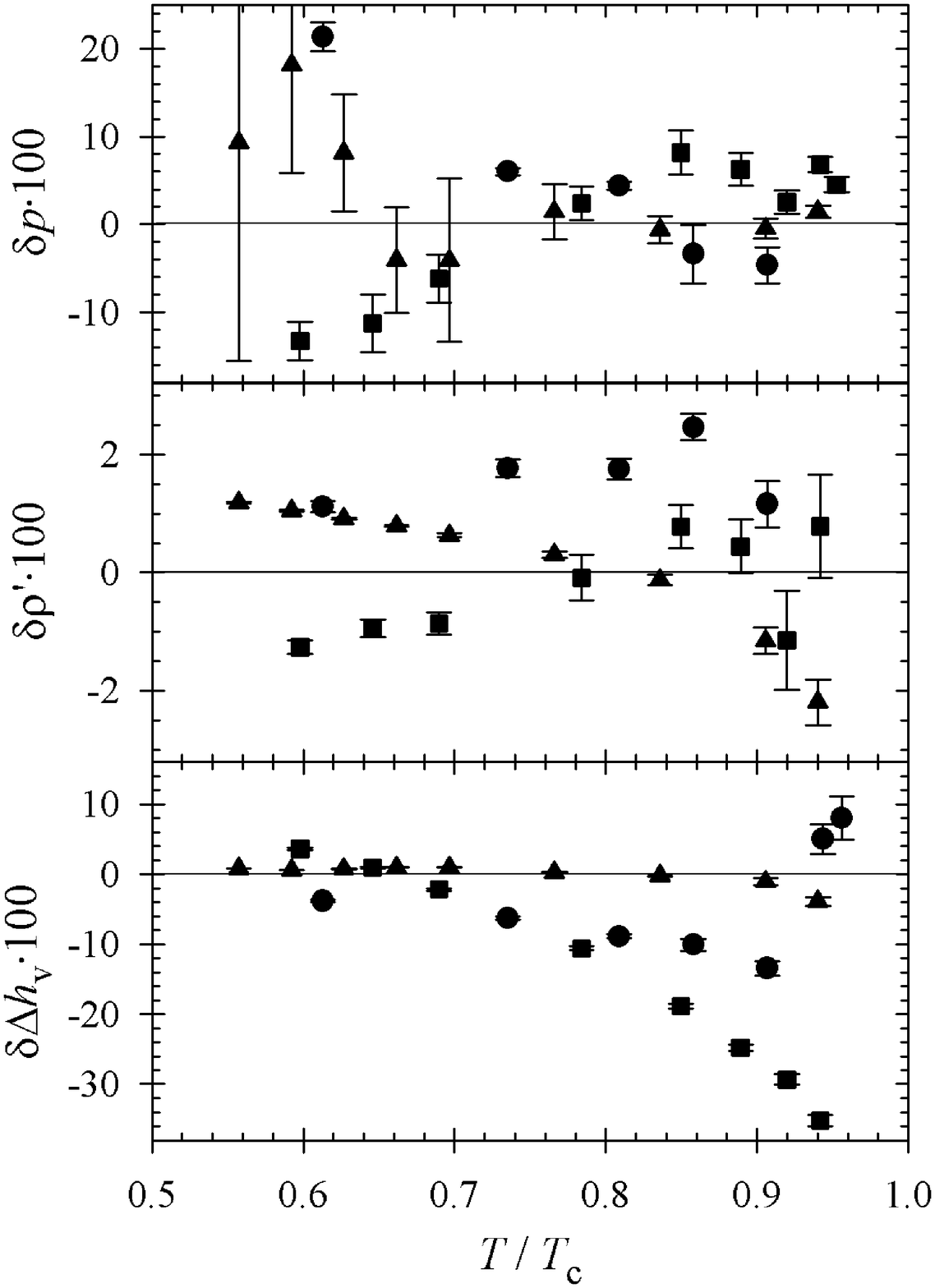}
\bigskip
\caption[Relative deviations of vapor-liquid equilibrium properties between simulation data and correlations of experimental data \cite{DIPPR2006} ($\delta z=(z_\mathrm{sim}-z_\mathrm{exp})/z_\mathrm{exp}$) for present models: {\Large $\bullet$}~formaldehyde, {\small $\blacksquare$}~hydrogen cyanide, {$\blacktriangle$}~sulfur dioxide. Top: vapor pressure, center: saturated liquid density, bottom: enthalpy of vaporization.]{\label{fig_vle3_dev}Eckl et al.}
\end{figure}

Figures~\ref{fig_vle3_rho} to \ref{fig_vle3_dhv} compare simulation results and experimental VLE data for form\-aldehyde, Figure~\ref{fig_vle3_dev} shows the relative deviations. Mean unsigned errors in vapor pressure, saturated liquid density, and enthalpy of vaporization are 4.3, 0.9, and 8.4\%, respectively.

Note that, in contrast to iso-butane or cyclohexane, the available experimental data base is very weak here. In fact, for vapor pressure only a single data set from the year 1935 \cite{Spence1935} is available. For saturated liquid density and enthalpy of vaporization, respectively, a single data point at 254~K is ``accepted'' by the DIPPR database \cite{DIPPR2006}. Thus, no further optimization of the molecular model was attempted although the desired quality seems not to be fully reached. Hermida-Ra\'on and R\'ios \cite{Hermida-Raon1998} published a molecular model based on QM calculations of formaldehyde dimers and trimers. They applied their model to liquid phase simulations but report no results on VLE properties. Thus, no comparison is presented here.

Dimethyl ether was modeled with three LJ sites in the present work, one for the oxygen atom and one for each methyl group. A dipole was located in the center of mass and oriented along the symmetry axis of the molecule, where the dipole moment was again taken from QM calculation. For an initial model, the same LJ parameters of the methyl groups were taken as for the iso-butane model, i.e. those by Ungerer et al. \cite{Ungerer2000}. An adjustment of the two LJ parameters of the oxygen site was sufficient to reach the desired quality. Alternative models for dimethyl ether are given in \cite{Jorgensen1981, Lin1995, Stubbs2004, Ketko2007}.

\begin{figure}[ht]
\includegraphics[scale=0.5]{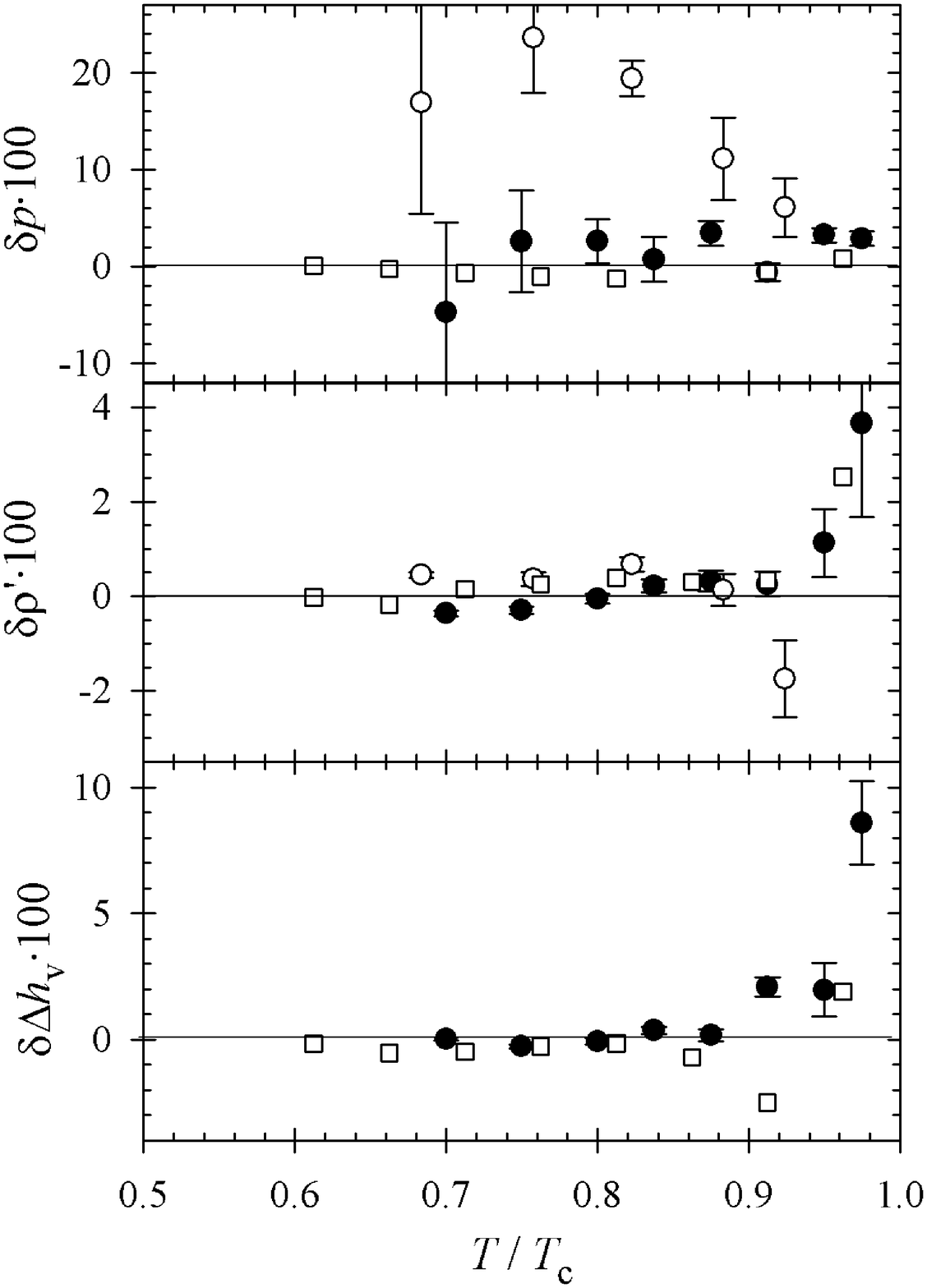}
\bigskip
\caption[Relative deviations of vapor-liquid equilibrium properties between simulation data and correlations of experimental data \cite{DIPPR2006} ($\delta z=(z_\mathrm{sim}-z_\mathrm{exp})/z_\mathrm{exp}$) for dimethyl ether: {\Large $\bullet$}~present model, {\Large $\circ$}~Stubbs et al. \cite{Stubbs2004}, {\small $\square$}~Ketko and Potoff \cite{Ketko2007}. Top: vapor pressure, center: saturated liquid density, bottom: enthalpy of vaporization.]{\label{fig_vle_dme_dev}Eckl et al.}
\end{figure}

The simulation results for the present model of dimethyl ether in comparison to the experimental VLE data are shown in Figures~\ref{fig_vle1_rho} to \ref{fig_vle1_dhv}. Figure~\ref{fig_vle_dme_dev} shows the relative deviations between simulation and experiment also including simulation results from Stubbs et al. \cite{Stubbs2004} and very recent results from Ketko and Potoff \cite{Ketko2007}. For dimethyl ether a good experimental data base is available for optimization of the molecular model. The simulation data of the present model is in very good agreement with the correlations of experimental data. The mean unsigned errors in vapor pressure, saturated liquid density, and enthalpy of vaporization are 2.6, 0.4, and 1.0\%, respectively.

For vapor pressure, both molecular models specifically adjusted to dimethyl ether, i.e. the present model and the model by Ketko and Potoff, yield better results than the transferable molecular model by Stubbs et al., while for saturated density all models perform similarly. For enthalpy of vaporization, the present model and the model by Ketko and Potoff also yield comparable results. Note that the parameters for the electrostatic interactions of the model by Ketko and Potoff were adjusted to experimental VLE data as well.

It can be summarized that the model by Ketko and Potoff \cite{Ketko2007} and the present dimethyl ether model are of similar quality and outperform the transferable model by Stubbs et al. \cite{Stubbs2004}. While Ketko and Potoff adjusted four LJ parameters and the point charge magnitudes for their electrostatic interactions, following the proposed modeling strategy an optimization of only two Lennard-Jones parameters was sufficient to reach the same quality.

\subsection{Sulfur Dioxide, Dimethyl Sulfide, and Thiophene}
For sulfur dioxide, a molecular model was published by Sokoli\'c et al. \cite{Sokolic1985, Sokolic1985a} which was optimized to total energy and pressure in the liquid state. It was recently reviewed by Ribeiro \cite{Ribeiro2006}. Alternatively, the commercial force field COMPASS \cite{Yang2000} reports parameters for sulfur dioxide.

For the present molecular model, the intermolecular interactions of sulfur dioxide were modeled with three LJ sites, i.e. one per atom, plus one dipole and one quadrupole. The electrostatic sites are located in the center of mass and parameterized according to the results of QM calculation. The four parameters of the LJ sites, i.e. $\sigma_\mathrm{S}$, $\varepsilon_\mathrm{S}$, $\sigma_\mathrm{O}$, and $\varepsilon_\mathrm{O}$, were adjusted to experimental VLE data. All parameters of the molecular model are given in Table~\ref{tab_Parameter}.

\begin{figure}[ht]
\includegraphics[scale=0.5]{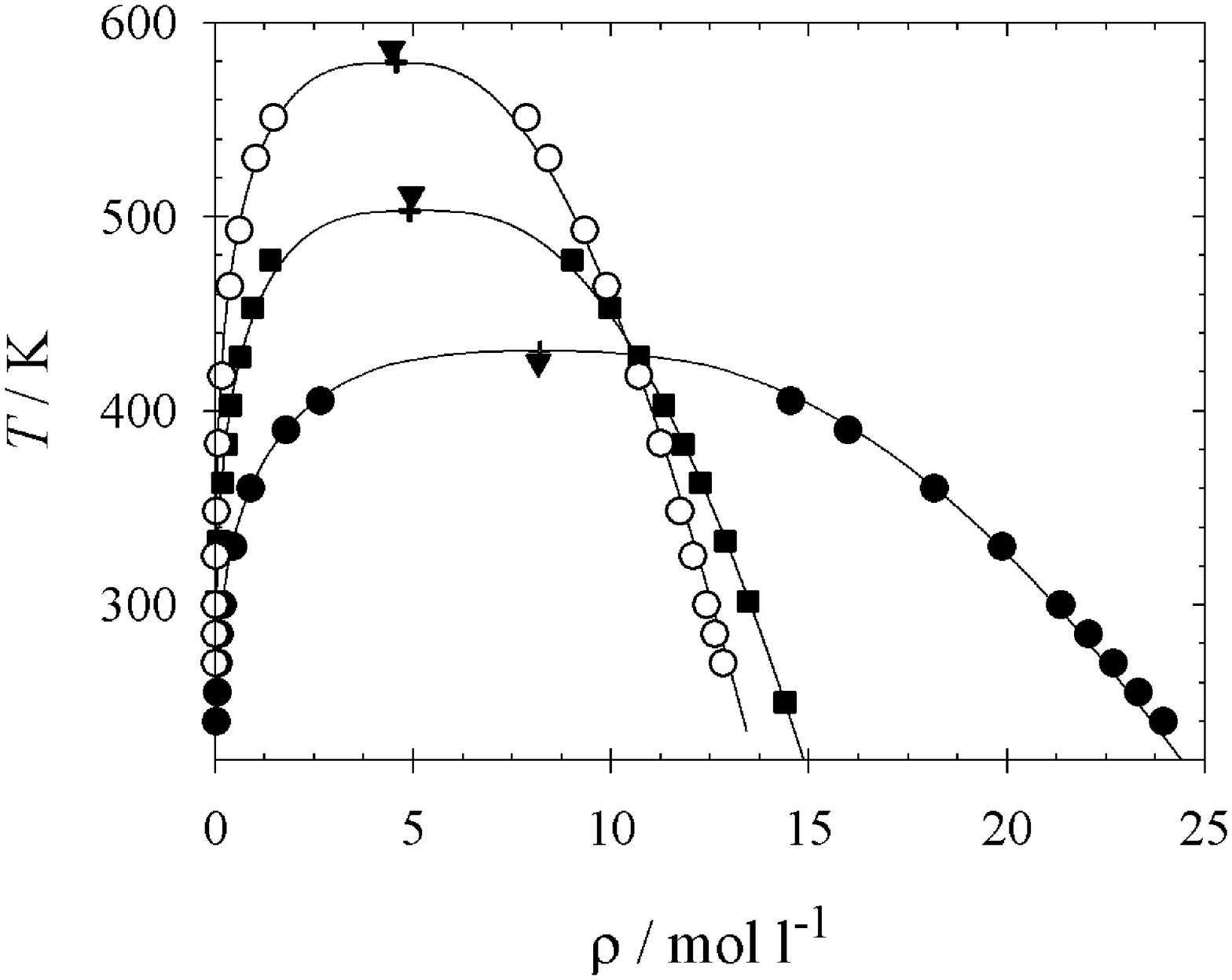}
\bigskip
\caption[Saturated densities. Present simulation data: {\Large $\bullet$}~sulfur dioxide, {\small $\blacksquare$}~dimethyl sulfide, {\Large $\circ$}~thiophene. {---}~correlations of experimental data \cite{DIPPR2006}, +~critical points derived from simulation, $\blacktriangledown$~experimental critical points, cf. Table~\ref{tab_critical}.]{\label{fig_vle2_rho}Eckl et al.} 
\end{figure}

\begin{figure}[ht]
\includegraphics[scale=0.5]{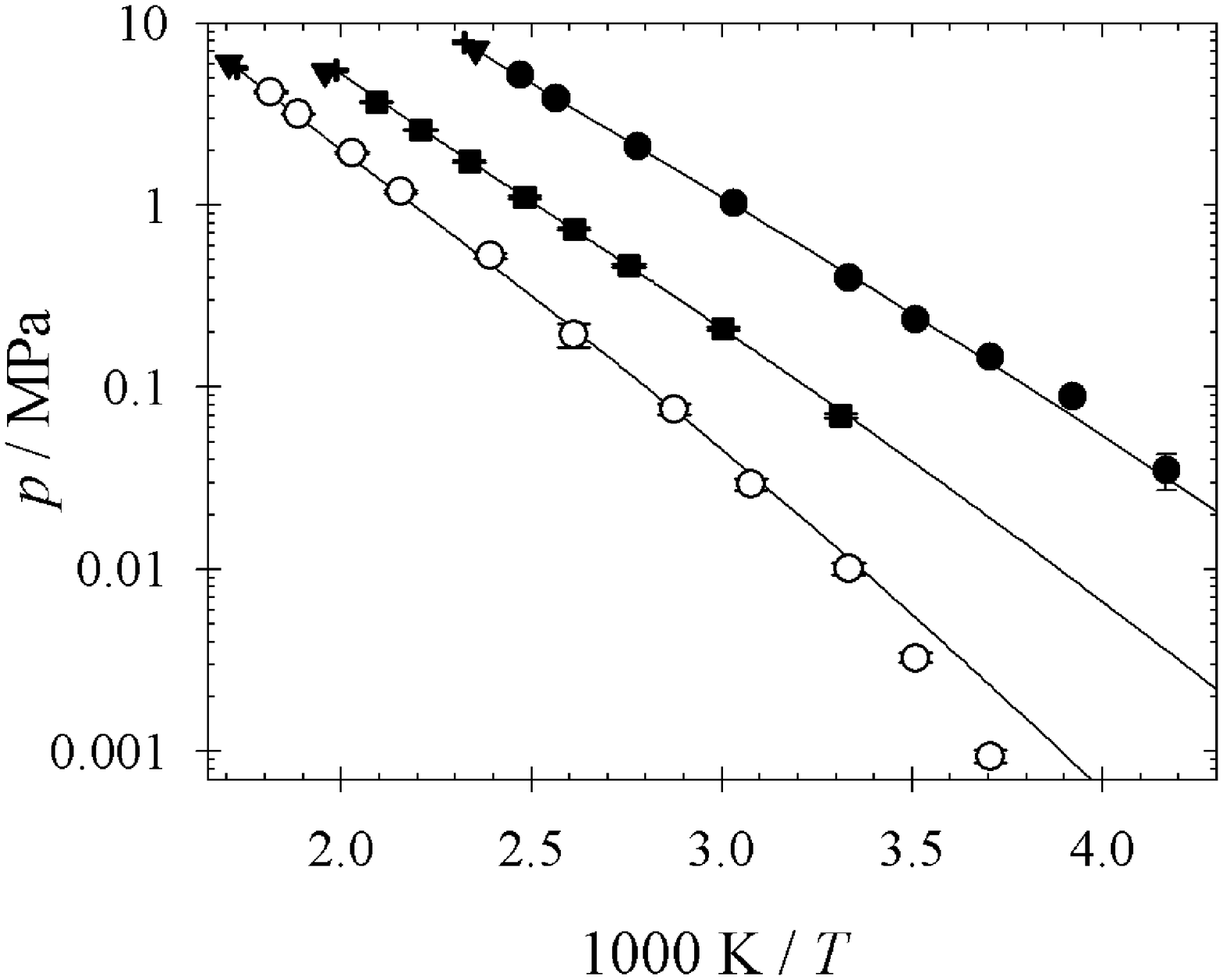}
\bigskip
\caption[Vapor pressure. Present simulation data: {\Large $\bullet$}~sulfur dioxide, {\small $\blacksquare$}~dimethyl sulfide, {\Large $\circ$}~thiophene. {---}~correlations of experimental data \cite{DIPPR2006}, +~critical points derived from simulation, $\blacktriangledown$~experimental critical points.]{\label{fig_vle2_p}Eckl et al.}
\end{figure}

\begin{figure}[ht]
\includegraphics[scale=0.5]{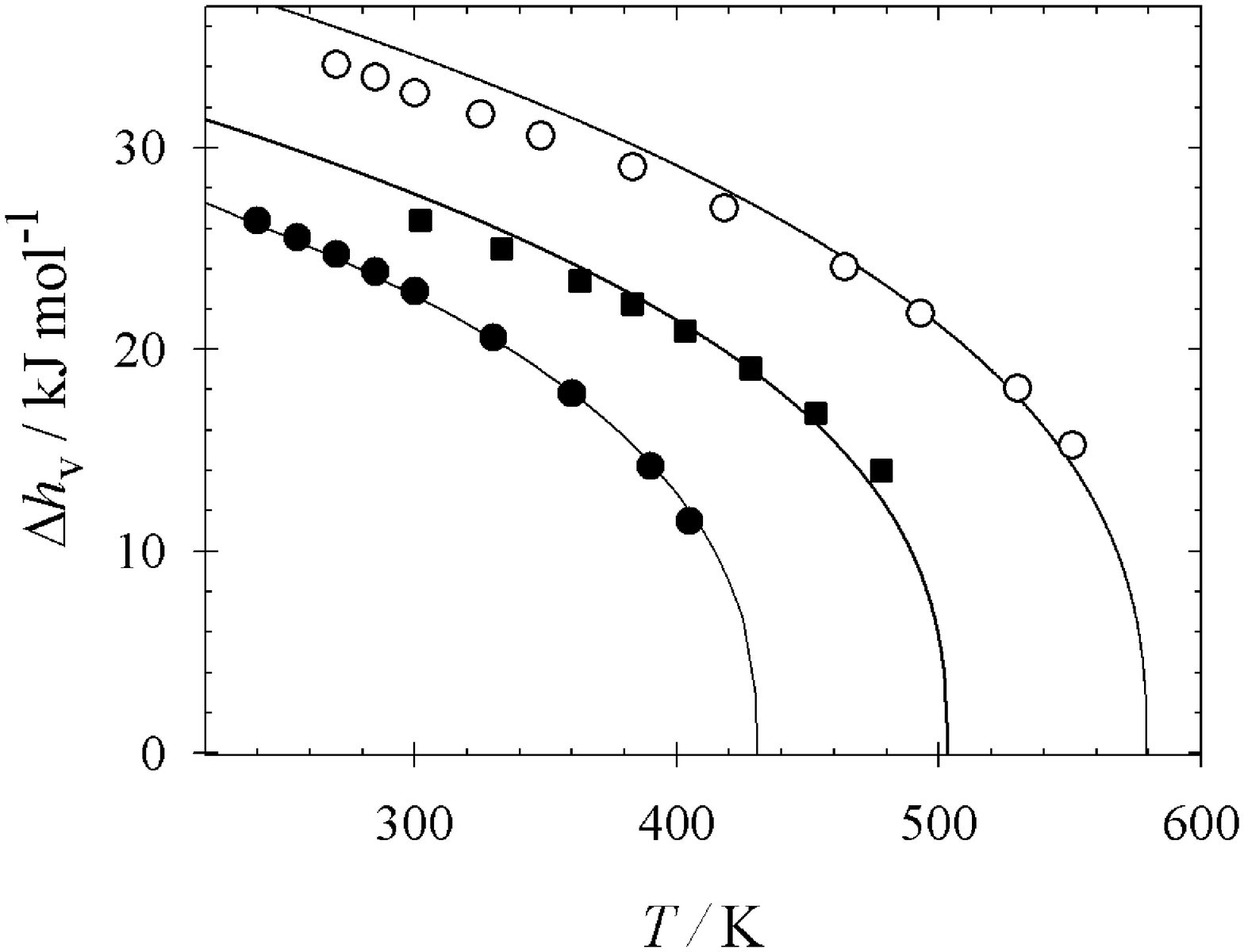}
\bigskip
\caption[Enthalpy of vaporization. Present simulation data: {\Large $\bullet$}~sulfur dioxide, {\small $\blacksquare$}~dimethyl sulfide, {\Large $\circ$}~thiophene. {---}~correlations of experimental data \cite{DIPPR2006}.]{\label{fig_vle2_dhv}Eckl et al.}
\end{figure}

Present simulation results for sulfur dioxide are compared to experimental VLE data in 
Figures~\ref{fig_vle2_rho} to \ref{fig_vle2_dhv}. Figure~\ref{fig_vle3_dev} shows the relative deviations between simulation and experiment for the present model, while no numerical VLE simulation data for sulfur dioxide from other authors was available to us. Mean unsigned errors in vapor pressure, saturated liquid density, and enthalpy of vaporization are 4.0, 0.9, and 1.6\%, respectively. The good experimental data base, with more than 60 individual experimental VLE data points, allows a thorough optimization of the molecular model to the desired quality.

Literature models for dimethyl sulfide are given in \cite{Jorgensen1986, Delhommelle2000, Lubna2005}. The present dimethyl sulfide model consists of three LJ sites, one for the sulfur atom and one for each methyl group. The electrostatic interactions are modeled by one dipole and two quadrupoles oriented perpendicularly to each other. This description of the electrostatics was chosen, as QM yields a charge distribution, which can not properly be described with a lower number of electrostatic sites. The LJ parameters of the methyl groups were assumed to be the same as for iso-butane and dimethyl oxide, while the parameters of the sulfur group were adjusted to experimental VLE data.

\begin{figure}[ht]
\includegraphics[scale=0.5]{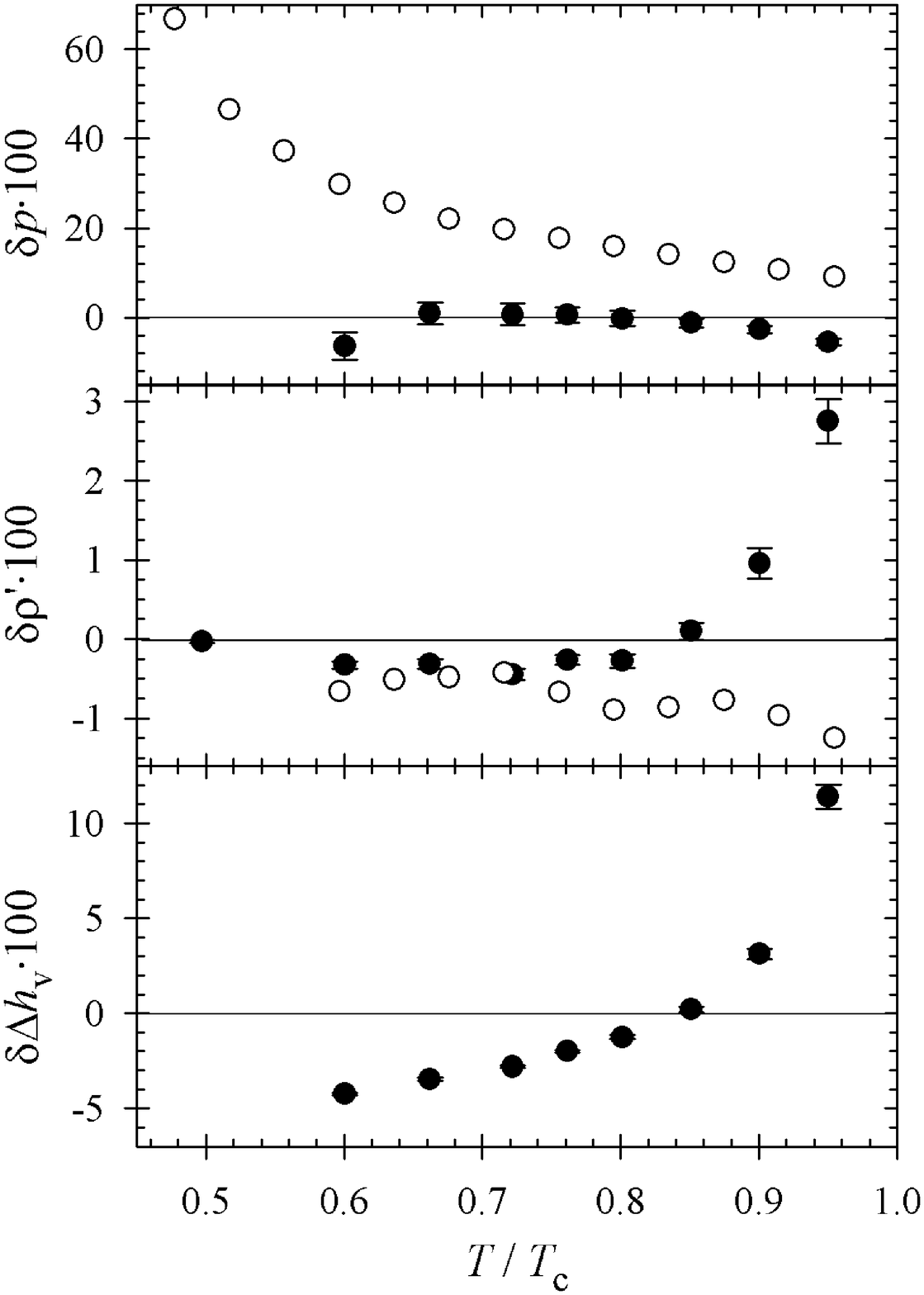}
\bigskip
\caption[Relative deviations of vapor-liquid equilibrium properties between simulation data and correlations of experimental data \cite{DIPPR2006} ($\delta z=(z_\mathrm{sim}-z_\mathrm{exp})/z_\mathrm{exp}$) for dimethyl sulfide: {\Large $\bullet$}~present  model, {\Large $\circ$}~Lubna et al. \cite{Lubna2005}. Top: vapor pressure, center: saturated liquid density, bottom: enthalpy of vaporization.]{\label{fig_vle_dms_dev}Eckl et al.}
\end{figure}

Figures~\ref{fig_vle2_rho} to \ref{fig_vle2_dhv} show the present simulation results for dimethyl sulfide in comparison to experimental VLE data. Figure~\ref{fig_vle_dms_dev} shows the relative deviations between simulation and experiment for the present model and for the model of Lubna et al. \cite{Lubna2005}. Note that simulation results on enthalpy of vaporization were not included in \cite{Lubna2005}. For the present model of dimethyl sulfide, mean unsigned errors in vapor pressure, saturated liquid density, and enthalpy of vaporization are 4.0, 0.7, and 3.8\%, respectively. Particularly the vapor pressure is better described than by the model of Lubna et al.

Also thiophene was described in the present work by a rigid model in its most stable conformation, as for cyclohexane. Five LJ sites, one for each of the four methylene groups and one for the sulfur atom, as well as one dipole and one quadrupole were used. The electrostatic parameters were directly passed on from QM. A total of four LJ parameters, i.e. $\sigma_\mathrm{CH2}$, $\varepsilon_\mathrm{CH2}$, $\sigma_\mathrm{S}$, and $\varepsilon_\mathrm{S}$, were adjusted to experimental VLE data. The optimized parameters are given in Table~\ref{tab_Parameter}. Alternative molecular models for thiophene can be found in the literature \cite{Lubna2005, Juarez-Guerra2006, Perez-Pellitero2007}

\begin{figure}[ht]
\includegraphics[scale=0.5]{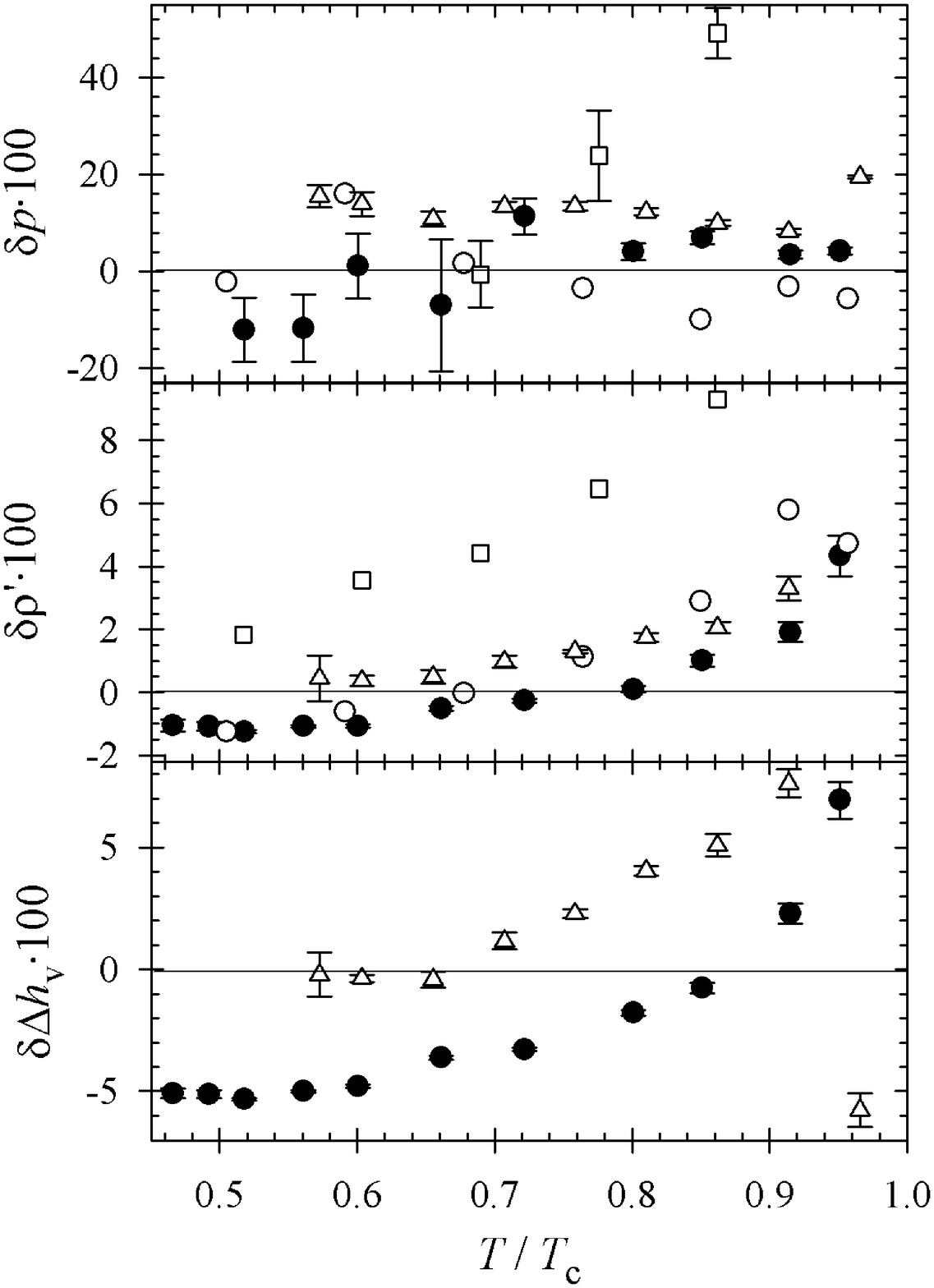}
\bigskip
\caption[Relative deviations of vapor-liquid equilibrium properties between simulation data and correlations of experimental data \cite{DIPPR2006} ($\delta z=(z_\mathrm{sim}-z_\mathrm{exp})/z_\mathrm{exp}$) for thiophene: {\Large $\bullet$}~present  model, {\Large $\circ$}~Lubna et al. \cite{Lubna2005}, {\small $\square$}~Ju\'{a}rez-Guerra et al. \cite{Juarez-Guerra2006}, $\triangle$~P\'{e}rez-Pellitero et al. \cite{Perez-Pellitero2007}. Top: vapor pressure, center: saturated liquid density, bottom: enthalpy of vaporization.]{\label{fig_vle_thiophen_dev}Eckl et al.}
\end{figure}

Figures~\ref{fig_vle2_rho} to \ref{fig_vle2_dhv} compare simulation results to experimental VLE data for thiophene, while the relative deviations are shown in Figure~\ref{fig_vle_thiophen_dev}. In the deviation plot also simulation results from Lubna et al. \cite{Lubna2005}, Ju\'{a}rez-Guerra et al. \cite{Juarez-Guerra2006}, and P\'{e}rez-Pellitero et al. \cite{Perez-Pellitero2007} are included. Mean unsigned errors of the present model in vapor pressure, saturated liquid density, and enthalpy of vaporization are 3.8, 1.2, and 3.2\%, respectively.

The thiophene model by Ju\'arez-Guerra et al. \cite{Juarez-Guerra2006} shows significant deviations in both vapor pressure and saturated liquid density while no simulation results for the enthalpy of vaporization were given by the authors. The anisotropic united atoms (AUA) potential by P\'erez-Pellitero et al. \cite{Perez-Pellitero2007} overpredicts the vapor pressure over the complete temperature range by up to 20\%. Simulation results for saturated liquid density and enthalpy of vaporization are in very good agreement with the DIPPR correlation for low temperatures but give noticeably higher values than the DIPPR correlation for temperatures above $0.7 T_\mathrm{c}$. The TraPPE force field by Lubna et al. \cite{Lubna2005} yields vapor pressure results that agree very well with the DIPPR correlation and the simulation results within their scatter. Saturated densities are higher than the correlation towards the critical point for all four molecular models. For enthalpy of vaporization no numerical data were given by Lubna et al.

It should be noted that for thiophene experimental vapor pressure data is available for temperatures up to around $0.9~T_\mathrm{c}$, experimental saturated liquid densities and enthalpies of vaporization are only available up to approximately $0.63~T_\mathrm{c}$. Thus, an assessment of the different molecular models regarding density and enthalpy of vaporization above $0.7~T_\mathrm{c}$ on the basis of the DIPPR correlations is questionable.

\subsection{Hydrogen Cyanide, Acetonitrile, and Nitromethane}
Hydrogen cyanide was modeled in the present work with two LJ sites, one for the methine group and one for the nitrogen atom. Electrostatic interactions were modeled by one dipole and one quadrupole oriented along the symmetry axis, where the parameters were passed on from QM. All four LJ parameters were adjusted to experimental VLE data. The optimized parameters can be found in Table~\ref{tab_Parameter}. For hydrogen cyanide no other molecular models were found in the literature.

Figures~\ref{fig_vle2_rho} to \ref{fig_vle2_dhv} compare simulation results and experimental VLE data for hydrogen cyanide. Figure~\ref{fig_vle3_dev} shows the relative deviations. Unfortunately, the DIPPR database \cite{DIPPR2006} contains no true experimental data on the enthalpy of vaporization for hydrogen cyanide. Consequently, in the optimization process, only minor attention was paid to the enthalpy of vaporization. Mean unsigned errors in vapor pressure, saturated liquid density, and enthalpy of vaporization are nominally 7.2, 1.0, and 12.2\%, respectively.

Several molecular models for acetonitrile are available in the literature. Jorgensen and Briggs \cite{Jorgensen1988}, Price et al. \cite{Price2001}, Gu\'ardia et al. \cite{Guardia2001}, and Nikitin and Lyubartsev \cite{Nikitin2007} present models that were optimized to the liquid density and enthalpy of vaporization at 293~K. Hloucha and Deiters \cite{Hloucha1997} give a polarizable molecular model for simulations in the liquid state while Hloucha et al. \cite{Hloucha2000} published a model that is based on \emph{ab initio} calculations. Finally, Wick et al. \cite{Wick2005} proposed an extension of their TraPPE force field that was optimized to VLE data to cover acetonitrile.

In the present work, acetonitrile was modeled using three LJ sites, one for the methyl group, one for the central carbon atom, and one for the nitrogen atom. The electrostatic interactions were modeled by one dipole and one quadrupole, located in the center of mass and parameterized strictly according to QM results. The parameters of the LJ sites of the methyl group and the nitrogen atom were adjusted to experimental VLE data. The LJ parameters of the central carbon atom were taken from unpublished work on carbon dioxide and excluded from optimization, only a very weak sensitivity of the VLE simulation results on these parameters was found. All parameters of the molecular model are given in Table~\ref{tab_Parameter}.

\begin{figure}[ht]
\includegraphics[scale=0.5]{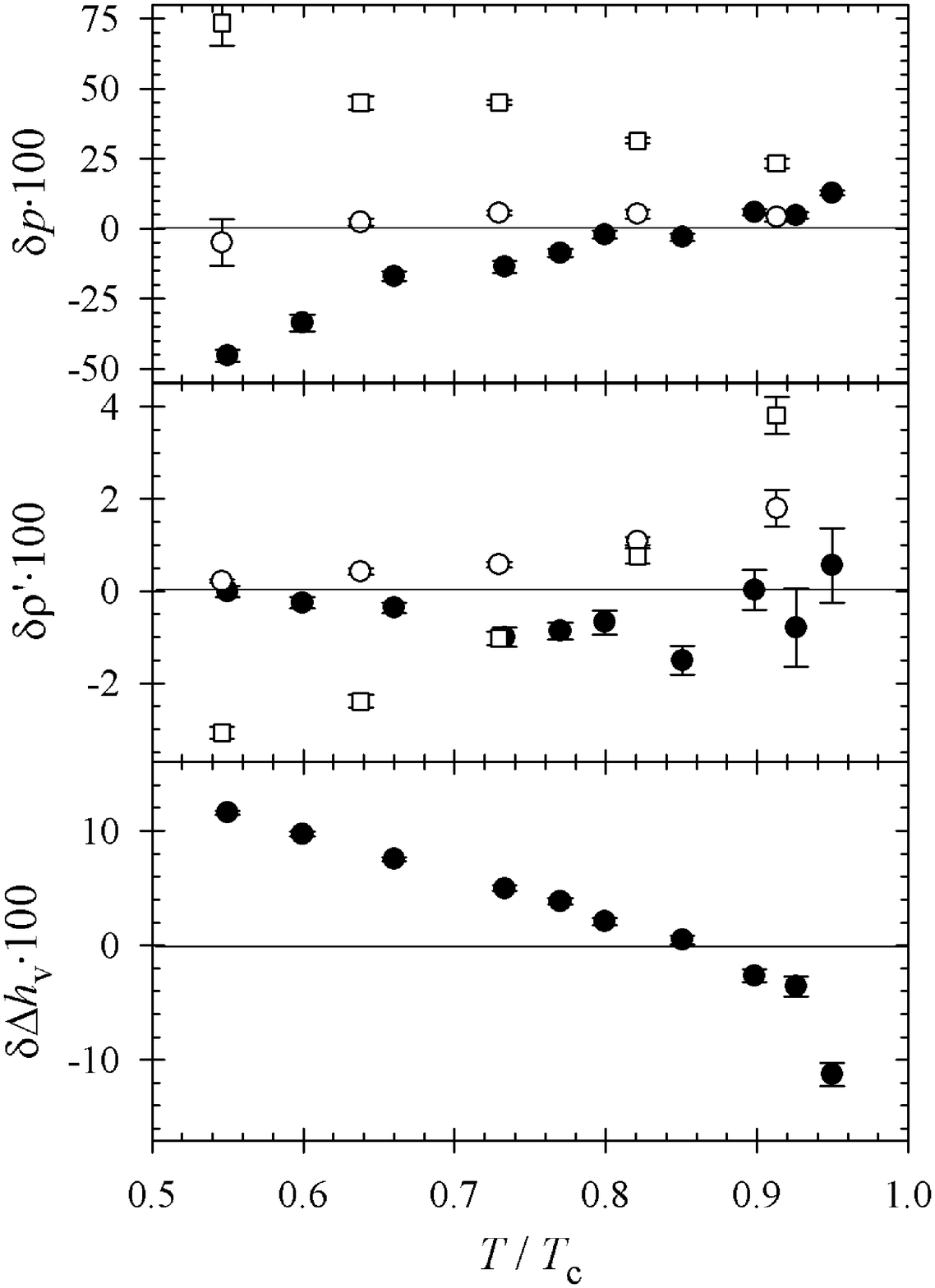}
\bigskip
\caption[Relative deviations of vapor-liquid equilibrium properties between simulation data and correlations of experimental data \cite{DIPPR2006} ($\delta z=(z_\mathrm{sim}-z_\mathrm{exp})/z_\mathrm{exp}$) for acetonitrile: {\Large $\bullet$}~present  model, {\Large $\circ$}~Wick et al. \cite{Wick2005}, {\small $\square$}~Jorgensen and Briggs \cite{Jorgensen1988, Wick2005}. Top: vapor pressure, center: saturated liquid density, bottom: enthalpy of vaporization.]{\label{fig_vle_acetonitril_dev}Eckl et al.}
\end{figure}

Simulation results for acetonitrile are compared to experimental VLE data in 
Figures~\ref{fig_vle3_rho} to \ref{fig_vle3_dhv}. The relative deviations between simulation and experiment are shown in Figure~\ref{fig_vle_acetonitril_dev} next to results of the OPLS-UA force field by Jorgensen and Briggs \cite{Jorgensen1988} and the TraPPE force field by Wick et al., which were reported for both models in \cite{Wick2005}. Despite the good experimental data base, the desired quality was not achieved by the present optimization. Only a fair description of the experimental VLE was reached. Mean unsigned errors in vapor pressure, saturated liquid density, and enthalpy of vaporization are 19.7, 0.9, and 5.4\%, respectively.

The large relative deviations in vapor pressure result from systematic underestimations at low temperatures. In this region also difficulties were encountered in the simulative calculation of the chemical potential in the liquid phase to determine the phase equilibrium. This was even the case when the more sophisticated gradual insertion method \cite{Vrabec2002a} was used, as described in the appendix in greater detail.

The OPLS-UA force field by Jorgensen and Briggs \cite{Jorgensen1988} significantly overestimates the vapor pressure of acetonitrile and shows deviations in the saturated liquid density up to 4\%, cf. Figure~\ref{fig_vle_acetonitril_dev}. Simulation results for vapor pressure obtained with the TraPPE force field by Wick et al. \cite{Wick2005} show a very good agreement with the DIPPR correlation, while the saturated density is slightly overpredicted towards the critical point. No comparison of the literature models for acetonitrile is possible regarding the enthalpy of vaporization due to the lack of numerical simulation results.

The present molecular model for nitromethane consists of four LJ sites, one for the methyl group, one for the nitrogen atom, and one for the two oxygen atoms each. The electrostatic interactions were modeled by one dipole and one quadrupole oriented along the symmetry axis, where the parameters were specified according to QM results. All six LJ parameters were adjusted to experimental VLE data. Alternative models are available in the literature \cite{Alper1999, Sorescu2001, Price2001, Wick2005}.

\begin{figure}[ht]
\includegraphics[scale=0.5]{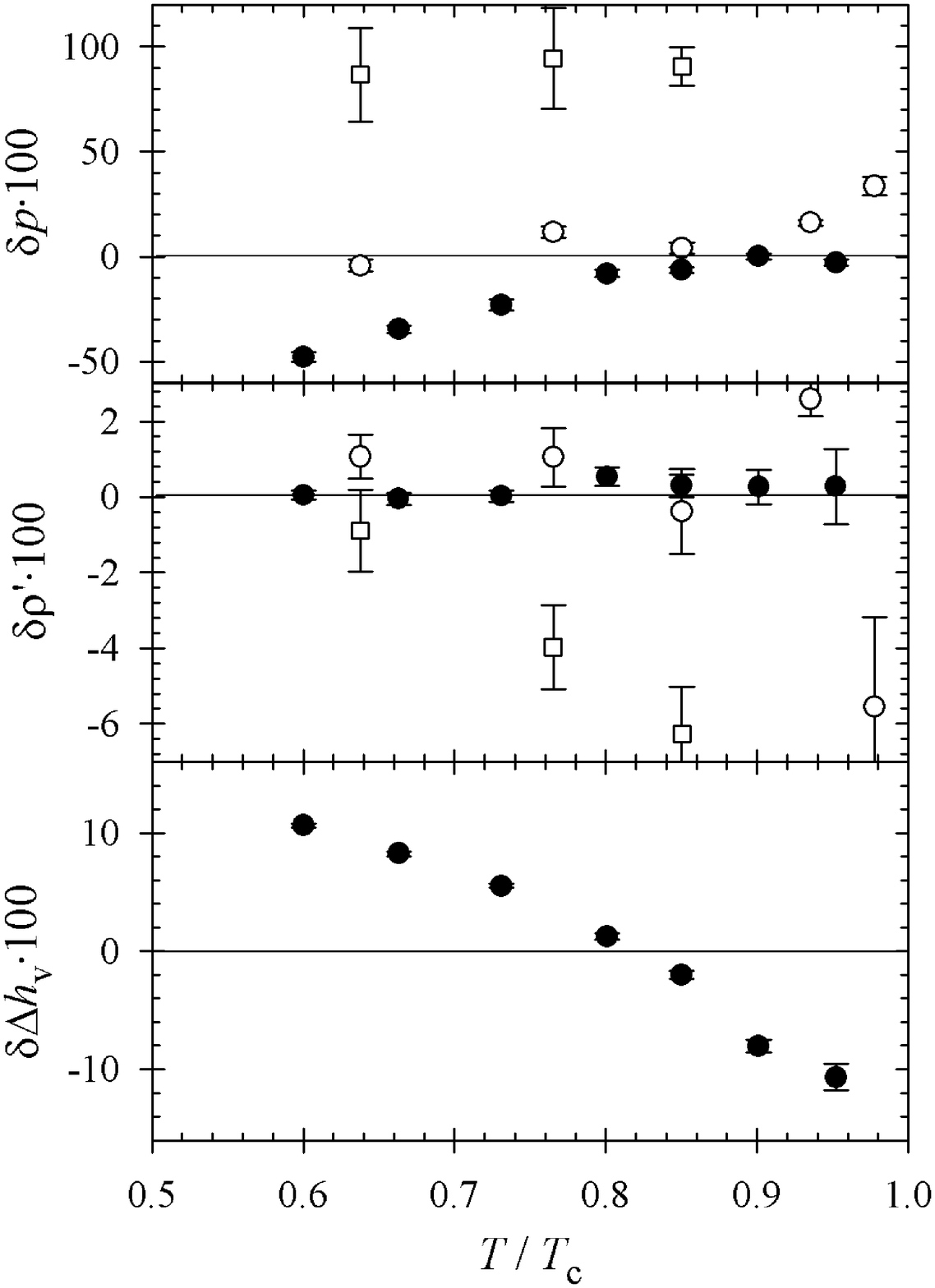}
\bigskip
\caption[Relative deviations of vapor-liquid equilibrium properties between simulation data and correlations of experimental data \cite{DIPPR2006} ($\delta z=(z_\mathrm{sim}-z_\mathrm{exp})/z_\mathrm{exp}$) for nitromethane: {\Large $\bullet$}~present  model, {\Large $\circ$}~Wick et al. \cite{Wick2005}, {\small $\square$}~Price et al. \cite{Price2001}. Top: vapor pressure, center: saturated liquid density, bottom: enthalpy of vaporization.]{\label{fig_vle_nitromethan_dev}Eckl et al.}
\end{figure}

Figures~\ref{fig_vle2_rho} to \ref{fig_vle2_dhv} compare the present simulation results for nitromethane with experimental VLE data. Figure~\ref{fig_vle_nitromethan_dev} shows the relative deviations obtained with the present model, the OPLS-AA force field by Price et al. \cite{Price2001}, and the TraPPE force field by Wick et al. \cite{Wick2005}. For the present model of nitromethane, mean unsigned errors in vapor pressure, saturated liquid density, and enthalpy of vaporization are 18.7, 0.2, and 7.0\%, respectively. Again, a systematic underprediction of the vapor pressure at low temperatures was found, leading to the high relative deviations, as for acetonitrile.

The OPLS-AA force field overpredicts the vapor pressure by about 80\% and underpredicts the saturated liquid density by up to 6\%. The TraPPE force field of Wick et al. that was optimized to experimental VLE data, yields very good results in vapor pressure for low temperatures while the result for the highest simulated temperature deviates by +34\% from the DIPPR correlation. Regarding saturated liquid density, strong scatter and large statistical uncertainties of the simulation results are observed for both literature models.

\section{Conclusion}
A strategy was proposed for the rapid development of molecular models for engineering applications. The strategy relies on results from QM \emph{ab initio} calculations to include physically sound molecular properties and to reduce the number of adjustable parameters. Dispersive and repulsive interactions were modeled by LJ sites. Thus, the LJ interaction sites were located according to atom positions obtained by QM energy minimization on Hartree-Fock level. For the parameterization of the electrostatic interactions, QM calculations were performed using the M\o{}ller-Plesset~2 level of theory and the COSMO method. The resulting electron density distribution was reduced to ideal point dipoles and ideal linear point quadrupoles located in the center of mass. The moments and orientations of the dipoles and quadrupoles were passed on to the molecular models without any modification.

A united-atom approach was used for methine, methylene, and methyl groups to reduce the total number of interaction sites. The parameters of the LJ sites were initially taken from similar sites of other molecular models and were subsequently optimized to reproduce experimental VLE data, i.e. vapor pressure, saturated liquid density, and enthalpy of vaporization. It was aimed to achieve deviations between simulation and experiment of below 5, 1, and 5\% in vapor pressure, saturated liquid density, and enthalpy of vaporization, respectively.

The new modeling strategy was successfully applied to ten molecules from different substances classes, i.e. iso-butane, cyclohexane, formaldehyde, dimethyl ether, sulfur dioxide, dimethyl sulfide, thiophene, hydrogen cyanide, acetonitrile, and nitromethane. Simulation results for the different substances agree well with correlations of experimental data taken from the DIPPR database \cite{DIPPR2006}, with noticeable deviations in the vapor pressure at low temperatures for acetonitrile and nitromethane.

For the two elongated molecules acetonitrile and nitromethane, the reduction of the electrostatic interactions to sites located in the center of mass seems to be an over-simplification, as the optimization of the molecular models was not fully successful. Also an adjustment of the electrostatic parameters, while keeping the ratio of the polar moments constant (not reported here in detail), did not yield significant improvements in the quality of the molecular models. A further study, e.g. using two or more spatially distributed electrostatic sites, is beyond the scope of this paper.

For all other molecules a significant improvement compared to available molecular models from the literature was achieved. Through an optimization to substance specific experimental VLE data, a very good description of the phase equilibria was obtained. Furthermore, in case of dimethyl ether it was shown that with the proposed modeling strategy it was possible to reach the same model quality by optimization of just two model parameters compared to five optimized parameters for the model by Ketko and Potoff \cite{Ketko2007}.

\section{Acknowledgment}
The authors gratefully acknowledge financial support by Deutsche Forschungsgemeinschaft, Schwerpunktprogramm 1155 "Molecular Modeling and Simulation in Process Engineering". The simulations were performed the national super computer NEC SX-8 at the High Performance Computing Center Stuttgart (HLRS) under the grant MMHBF and on the HP XC6000 super computer at the Steinbuch Centre for Computing, Karlsruhe under the grant MMSTP.

The authors want to thank Inga Utz for her modeling work on the sulfur and nitrogen containing compounds.

\section{Appendix}
\label{Appendix}
The Grand Equilibrium method \cite{Vrabec2002} was used to calculate VLE data at seven to thirteen temperatures from 50 to 97\% of the critical temperature during the optimization process.
For the liquid, molecular dynamics simulations were performed in the isobaric-isothermal ($NpT$) ensemble using isokinetic velocity scaling \cite{Allen1987} and Anderson's barostat \cite{Anderson1980}. There, the number of molecules was $864$ throughout and the time step was $1$ to $3$~fs depending on the molecular weight and the magnitude of the intermolecular interactions. The initial configuration was a face centered cubic lattice, the fluid was equilibrated over $25~000$ time steps with the first $5~000$ time steps in the canonical ($NVT$) ensemble. The production run time span was $150~000$ to $200~000$ time steps with a membrane mass of $10^9$ kg/m$^4$. Widom's insertion method \cite{Widom1963} was used to calculate the chemical potential by inserting up to $4~000$ test molecules every production time step.

In cases where Widom's insertion method yielded large statistical uncertainties for the chemical potential, i.e. at high densities for strongly interacting molecules, Monte Carlo simulations were performed in the $NpT$ ensemble for the liquid. Then, the chemical potential was calculated by the gradual insertion method \cite{Nezbeda1991, Vrabec2002a}. The number of molecules was $500$. Starting from a face centered cubic lattice, $15~000$ Monte Carlo cycles were performed for equilibration and $50~000$ for production, each cycle containing $500$ translation moves, $500$ rotation moves, and $1$ volume move. Every $50$ cycles, $5000$ fluctuating state change moves, $5000$ fluctuating particle translation/rotation moves, and $25000$ biased particle translation/rotation moves were performed, to determine the chemical potential. These computationally demanding simulations yield the chemical potential in dense and strong interacting liquids with high accuracy, leading to reasonable uncertainties in the VLE.

For the corresponding vapor, Monte Carlo simulations in the pseudo-$\mu VT$ ensemble were performed. The simulation volume was adjusted to lead to an average number of $500$ molecules in the vapor phase. After $1~000$ initial $NVT$ Monte Carlo cycles, starting from a face centered cubic lattice, $10~000$ equilibration cycles in the pseudo-$\mu VT$ ensemble were performed. The length of the production run was $50~000$ cycles. One cycle is defined here to be a number of attempts to displace and rotate molecules equal to the actual number of molecules plus three insertion and three deletion attempts.

The cut-off radius was set to $17.5$~\r{A} throughout and a center of mass cut-off scheme was employed. Lennard-Jones long-range interactions beyond the cut-off radius were corrected employing angle averaging as proposed by Lustig \cite{Lustig1988}. Electrostatic interactions were approximated by a resulting molecular dipole and corrected using the reaction field method \cite{Allen1987}. Statistical uncertainties in the simulated values were estimated by a block averaging method \cite{Flyvbjerg1989}.



\newpage
\begin{landscape}
\begin{table}[ht]
\noindent
\caption{Parameters of the new molecular models. Lennard-Jones interaction sites are denoted by the modeled atoms (in boldface) with an additional bonding partner if necessary. Electrostatic interaction sites are denoted by dipole or quadrupole, respectively. Coordinates are given with respect to the center of mass in a principal axes system. Orientations of the electrostatic sites are defined in standard Euler angles, where $\varphi$ is the azimuthal angle with respect to the $x-z$ plane and $\theta$ is the inclination angle with respect to the $z$ axis.}
\label{tab_Parameter}

\medskip
\begin{center}
\begin{tabular}{lccccccccc} \hline\hline
Interaction        & $x$      & $y$      & $z$      & $\sigma$ & $\varepsilon/k_\mathrm{B}$
                                                                         & $\theta$ & $\varphi$ & $\mu$    & $Q$      \\
Site               & \r{A}    & \r{A}    & \r{A}    & \r{A}    & K       & $\deg$   & $\deg$    & D        & B        \\ \hline
\multicolumn{10}{l}{Iso-butane} \\
\textbf{CH}        & \my0     & \my0     & \.0.8179 & 3.360    & \051.00 & ---      & ---       & ---      & ---      \\
\textbf{CH$_3$}(1) & \.1.7302 & \my0     & -0.1893  & 3.607    & 120.15  & ---      & ---       & ---      & ---      \\
\textbf{CH$_3$}(2) & -0.8651  & \.1.4984 & -0.1893  & 3.607    & 120.15  & ---      & ---       & ---      & ---      \\
\textbf{CH$_3$}(3) & -0.8651  & -1.4984  & -0.1893  & 3.607    & 120.15  & ---      & ---       & ---      & ---      \\
Dipole             & \my0     & \my0     & \my0     & ---      & ---     & \0\00    & \0\00     & \.0.1347 & ---      \\
Quadrupole         & \my0     & \my0     & \my0     & ---      & ---     & \0\00    & \0\00     & ---      & \.0.7236 \\ \hline
\multicolumn{10}{l}{Cyclohexane} \\
\textbf{CH$_2$}(1) & \.0.0210 & -0.3118  & \.1.8052 & 3.497    & \087.39 & ---      & ---       & ---      & ---      \\
\textbf{CH$_2$}(2) & \.1.5318 & \.0.2989 & \.0.8863 & 3.497    & \087.39 & ---      & ---       & ---      & ---      \\
\textbf{CH$_2$}(3) & -1.5528  & \.0.2983 & \.0.9986 & 3.497    & \087.39 & ---      & ---       & ---      & ---      \\
\textbf{CH$_2$}(4) & \.1.5318 & -0.2989  & -0.8863  & 3.497    & \087.39 & ---      & ---       & ---      & ---      \\
\textbf{CH$_2$}(5) & -1.5528  & -0.2983  & -0.9986  & 3.497    & \087.39 & ---      & ---       & ---      & ---      \\
\textbf{CH$_2$}(6) & \.0.0210 & \.0.3118 & -1.8052  & 3.497    & \087.39 & ---      & ---       & ---      & ---      \\
Quadrupole         & \my0     & \my0     & \my0     & ---      & ---     & \090     & \090      & ---      & \.0.8179 \\ \hline
\multicolumn{10}{l}{Formaldehyde} \\
\textbf{O}         & \my0     & \my0     & \.0.6721 & 3.010    & 112.61  & ---      & ---       & ---      & ---      \\
\textbf{CH$_2$}    & \my0     & \my0     & -0.7682  & 3.422    & \077.42 & ---      & ---       & ---      & ---      \\
Dipole             & \my0     & \my0     & \.0.0480 & ---      & ---     & 180      & \0\00     & \.2.6668 & ---      \\ \hline
\multicolumn{10}{r}\emph{continued on next page}
\end{tabular}
\end{center}
\end{table}

\begin{table}[ht]
\begin{center}
\begin{tabular}{lccccccccc}
\multicolumn{10}{l}\emph{continued from previous page} \\
\hline
Interaction        & $x$      & $y$      & $z$      & $\sigma$ & $\varepsilon/k_\mathrm{B}$
                                                                         & $\theta$ & $\varphi$ & $\mu$    & $Q$     \\
Site               & \r{A}    & \r{A}    & \r{A}    & \r{A}    & K       & $\deg$   & $\deg$    & D        & B       \\ \hline
\multicolumn{10}{l}{Dimethyl Ether} \\
\textbf{O}         & \my0     & \my0     & \.0.6427 & 2.727    & \089.57 & ---      & ---       & ---      & ---      \\
\textbf{CH$_3$}(1) & \.1.4041 & \my0     & -0.3086  & 3.607    & 120.15  & ---      & ---       & ---      & ---      \\
\textbf{CH$_3$}(2) & -1.4041  & \my0     & -0.3086  & 3.607    & 120.15  & ---      & ---       & ---      & ---      \\
Dipole             & \my0     & \my0     & \my0     & ---      & ---     & 180      & \0\00     & \.1.7040 & ---      \\ \hline
\multicolumn{10}{l}{Sulfur Dioxide} \\
\textbf{S}         & \my0     & \my0     & \.0.3757 & 3.312    & 139.23  & ---      & ---       & ---      & ---      \\
\textbf{O}(1)      & \.1.2790 & \my0     & -0.3653  & 3.106    & \043.18 & ---      & ---       & ---      & ---      \\
\textbf{O}(2)      & -1.2790  & \my0     & -0.3653  & 3.106    & \043.18 & ---      & ---       & ---      & ---      \\
Dipole             & \my0     & \my0     & \my0     & ---      & ---     & \0\00    & \0\00     & \.1.9980 & ---      \\
Quadrupole         & \my0     & \my0     & \my0     & ---      & ---     & \090     & \0\00     & ---      & -5.3340  \\ \hline
\multicolumn{10}{l}{Dimethyl Sulfide} \\
\textbf{S}         & \my0     & \my0     & \.0.2819 & 3.398    & 207.57  & ---      & ---       & ---      & ---      \\
\textbf{CH$_3$}(1) & \.1.1583 & \my0     & -0.3868  & 3.607    & 120.15  & ---      & ---       & ---      & ---      \\
\textbf{CH$_3$}(2) & -1.1583  & \my0     & -0.3868  & 3.607    & 120.15  & ---      & ---       & ---      & ---      \\
Dipole             & \my0     & \my0     & \my0     & ---      & ---     & \090     & 180       & \.2.3610 & ---      \\
Quadrupole         & \my0     & \my0     & \my0     & ---      & ---     & \090     & \0\00     & ---      & \.3.0740 \\
Quadrupole         & \my0     & \my0     & \my0     & ---      & ---     & \0\00    & \0\00     & ---      & \.2.7600 \\ \hline
\multicolumn{10}{l}{Thiophene} \\
\textbf{S}         & \my0     & \my0     & \.0.7359 & 4.292    & \095.65 & ---       & ---       & ---      & ---      \\
\textbf{CH}(1)-S   & \.1.2513 & \my0     & \.0.2067 & 3.590    & \048.49 & ---       & ---       & ---      & ---      \\
\textbf{CH}(2)-S   & -1.2513  & \my0     & \.0.2067 & 3.590    & \048.49 & ---       & ---       & ---      & ---      \\
\textbf{CH}(3)-CH  & \00.7734 & \my0     & -1.2643  & 3.590    & \048.49 & ---       & ---       & ---      & ---      \\
\textbf{CH}(4)-CH  & -0.7734  & \my0     & -1.2643  & 3.590    & \048.49 & ---       & ---       & ---      & ---      \\
Dipole             & \my0     & \my0     & \my0     & ---      & ---     & 180       & \0\00     & \.1.8120 & ---      \\
Quadrupole         & \my0     & \my0     & \my0     & ---      & ---     & \090      & \0\00     & ---      & \.6.5389 \\ \hline
\multicolumn{10}{r}\emph{continued on next page}
\end{tabular}
\end{center}
\end{table}

\begin{table}[ht]
\begin{center}
\begin{tabular}{lccccccccc}
\multicolumn{10}{l}\emph{continued from previous page} \\
\hline
Interaction        & $x$      & $y$      & $z$      & $\sigma$ & $\varepsilon/k_\mathrm{B}$
                                                                         & $\theta$ & $\varphi$ & $\mu$    & $Q$      \\
Site               & \r{A}    & \r{A}    & \r{A}    & \r{A}    & K       & $\deg$   & $\deg$    & D        & B        \\ \hline
\multicolumn{10}{l}{Hydrogen Cyanide} \\
\textbf{N}         & \my0     & \my0     & \.0.6380 & 3.233    & \039.69 & ---      & ---       & ---      & ---      \\
\textbf{CH}        & \my0     & \my0     & -0.9671  & 3.445    & 102.44  & ---      & ---       & ---      & ---      \\
Dipole             & \my0     & \my0     & \.0.0589 & ---      & ---     & 180      & \0\00     & \.3.4084 & ---      \\
Quadrupole         & \my0     & \my0     & \.0.0589 & ---      & ---     & \0\00    & \0\00     & ---      & \.2.1800 \\ \hline
\multicolumn{10}{l}{Acetonitrile} \\
\textbf{N}         & \my0     & \my0     & \.1.1507 & 3.368    & \053.00 & ---      & ---       & ---      & ---      \\
\textbf{C}         & \my0     & \my0     & \.0.0507 & 2.810    & \010.64 & ---      & ---       & ---      & ---      \\
\textbf{CH$_3$}    & \my0     & \my0     & -1.2868  & 3.835    & 163.04  & ---      & ---       & ---      & ---      \\
Dipole             & \my0     & \my0     & \my0     & ---      & ---     & 180      & \0\00     & \.4.1186 & ---      \\
Quadrupole         & \my0     & \my0     & \my0     & ---      & ---     & \0\00    & \0\00     & ---      & -3.1373  \\ \hline
\multicolumn{10}{l}{Nitromethane} \\
\textbf{N}         & \my0     & \my0     & \.0.2199 & 3.321    & \034.90 & ---      & ---       & ---      & ---      \\
\textbf{O}(1)      & \.1.1045 & \my0     & \.0.7858 & 3.060    & \045.17 & ---      & ---       & ---      & ---      \\
\textbf{O}(2)      & -1.1045  & \my0     & \.0.7858 & 3.060    & \045.17 & ---      & ---       & ---      & ---      \\
\textbf{CH$_3$}    & \my0     & \my0     & -1.5135  & 3.501    & 158.79  & ---      & ---       & ---      & ---      \\
Dipole             & \my0     & \my0     & \.0.2535 & ---      & ---     & 180      & \0\00     & \.3.9901 & ---      \\
Quadrupole         & \my0     & \my0     & \.0.2535 & ---      & ---     & \090     & \0\00     & ---      & -4.7903  \\ \hline\hline
\end{tabular}
\end{center}
\end{table}
\end{landscape}

\newpage
\begin{table}[ht]
\noindent
\caption{Critical properties: present simulation results compared to recommended experimental data. The numbers in parentheses indicate the experimental uncertainty in the last digits.}
\label{tab_critical}

\medskip
\begin{center}
\begin{tabular}{lccccccc} \hline\hline
                 & $T_c^{\mathrm{sim}}$
                       & $T_c^{\mathrm{exp}}$
                                    & $\rho_c^{\mathrm{sim}}$
                                            & $\rho_c^{\mathrm{exp}}$
                                                         & $p_c^{\mathrm{sim}}$
                                                                & $p_c^{\mathrm{exp}}$
                                                                             & Ref. \\
                 & K   & K          & mol/l & mol/l      & MPa  & MPa        &      \\ \hline
Iso-butane       & 407 & 407.8(5)   & 3.87  & 3.86(5)\0  & 3.65 & 3.64(5)\0  & \cite{Daubert1996} \\
Cyclohexane      & 556 & 553.8(2)   & 3.26  & 3.25(2)\0  & 4.23 & 4.08(3)\0  & \cite{Daubert1996} \\
Formaldehyde     & 406 & 408\phantom{.0(0)}
                                    & 8.38  & 8.70\phantom{(00)}
                                                         & 5.95 & 6.59\phantom{(00)}
                                                                             & \cite{Reid1977} \\
Dimethyl Ether   & 403 & 400.2(1)   & 5.99  & 5.95(2)\0  & 5.69 & 5.34(5)\0  & \cite{Kudchadker2001} \\
Sulfur Dioxide   & 425 & 430.7(1)   & 8.15  & 8.2\0(8)\0 & 7.22 & 7.9\0(4)\0 & \cite{Mathews1972} \\
Dimethyl Sulfide & 511 & 503\.\0(1) & 4.95  & 4.91(8)\0  & 5.46 & 5.53(10)   & \cite{Tsonopoulos2001} \\
Hydrogen Cyanide & 448 & 457\.\0(1) & 7.89  & 7.4\0(1)\0 & 4.69 & 5.4\0(1)\0 & \cite{Marsh2006} \\
Acetonitrile     & 540 & 545.5(1)   & 6.04  & 5.8\0(1)\0 & 4.95 & 4.85(3)\0  & \cite{Marsh2006} \\
Nitromethane     & 587 & 588\.\0(1) & 5.87  & 5.8\0(0)\0 & 5.98 & 6.3\0(1)\0 & \cite{Kudchadker1968} \\ \hline\hline
\end{tabular}
\end{center}
\end{table}

\clearpage

\listoffigures

\newpage

\newpage

\newpage

\newpage

\newpage

\newpage

\newpage


\newpage

\newpage

\newpage

\newpage

\newpage

\newpage

\newpage

\newpage

\newpage

\newpage

\newpage

\noindent
{\large \bf Supplementary Material}

\begin{table}[H]
\noindent
\caption{Vapor-liquid equilibria of the modeled substances: present simulation results compared to experimental data \cite{DIPPR2006} for vapor pressure, saturated densities and heat of vaporization. The number in parentheses indicates the statistical uncertainty in the last digit.}
\label{tab_vle}

\medskip
\begin{center}
\begin{tabular}{c|ccccccc} \hline\hline
$T$ & $p_\sigma^{\mathrm{sim}}$
              & $p_\sigma^{\mathrm{exp}}$
                      & $\rho'_{\mathrm{sim}}$
                                  & $\rho'_{\mathrm{exp}}$
                                           & $\rho''_{\mathrm{sim}}$
                                                       & $\Delta h_\mathrm{v}^\mathrm{sim}$
                                                                    & $\Delta h_\mathrm{v}^\mathrm{exp}$ \\
K   & MPa     & MPa   & mol/l     & mol/l  & mol/l     & kJ/mol     & kJ/mol  \\ \hline
\multicolumn{8}{l}{Iso-butane} \\ \hline
150 &         &       & 12.15(1)  & 12.149 &           &            &         \\
200 &         &       & 11.28(1)  & 11.325 &           &            &         \\
270 & 0.14(1) & 0.140 & \09.98(1) & 10.050 & 0.067(6)  & 20.80(3)   & 20.811  \\
300 & 0.37(2) & 0.370 & \09.37(1) & \09.420 & 0.164(7)  & 19.10(3)   & 19.006  \\
320 & 0.59(2) & 0.633 & \08.88(2) & \08.955 & 0.254(8)  & 17.80(4)   & 17.591  \\
335 & 0.87(2) & 0.906 & \08.54(2) & \08.575 & 0.37\0(1) & 16.72(4)   & 16.375  \\
350 & 1.30(3) & 1.258 & \08.10(3) & \08.159 & 0.58\0(1) & 15.16(6)   & 14.980  \\
365 & 1.74(3) & 1.701 & \07.69(3) & \07.688 & 0.79\0(1) & 13.76(6)   & 13.325  \\
380 & 2.21(3) & 2.252 & \07.05(5) & \07.113 & 1.03\0(2) & 11.89(9)   & 11.240  \\
390 & 2.77(3) & 2.689 & \06.59(9) & \06.609 & 1.44\0(2) & \09.9\0(1) & \09.408 \\ \hline
\multicolumn{8}{l}{Cyclohexane} \\ \hline
300 &         &       & 9.13(0) & 9.174 &           &           &        \\
330 & 0.02(1) & 0.046 & 8.80(0) & 8.831 & 0.013(3)  & 31.49(1)  & 31.339 \\
360 & 0.12(1) & 0.121 & 8.45(0) & 8.476 & 0.041(4)  & 29.89(1)  & 29.589 \\
390 & 0.27(2) & 0.267 & 8.09(1) & 8.101 & 0.088(5)  & 28.19(2)  & 27.646 \\
415 & 0.51(2) & 0.467 & 7.77(1) & 7.767 & 0.165(7)  & 26.50(3)  & 25.828 \\
440 & 0.75(2) & 0.763 & 7.41(1) & 7.404 & 0.240(7)  & 24.78(3)  & 23.758 \\
460 & 1.08(2) & 1.085 & 7.12(1) & 7.083 & 0.343(6)  & 23.18(4)  & 21.853 \\
480 & 1.48(2) & 1.496 & 6.76(3) & 6.723 & 0.478(7)  & 21.27(5)  & 19.647 \\
500 & 2.01(2) & 2.012 & 6.38(2) & 6.305 & 0.68\0(1) & 18.99(7)  & 17.027 \\
515 & 2.44(2) & 2.478 & 5.97(3) & 5.935 & 0.85\0(1) & 16.92(9)  & 14.675 \\
535 & 3.29(2) & 3.223 & 5.44(6) & 5.280 & 1.31\0(2) & 13.1\0(1) & 10.566 \\ \hline
\multicolumn{8}{l}{Formaldehyde} \\ \hline
250 & 0.103(1)  & 0.085 & 27.81(3)  & 27.50 & 0.053(1)   & 22.41(4)  & 23.30  \\
300 & 0.582(2)  & 0.549 & 24.83(4)  & 24.40 & 0.273(3)   & 19.49(5)  & 20.80  \\
330 & 1.327(6)  & 1.271 & 22.69(4)  & 22.30 & 0.617(7)   & 17.22(5)  & 18.90  \\
350 & 1.99\0(7) & 2.058 & 21.11(5)  & 20.60 & 0.95\0(7)  & 15.6\0(2) & 17.30  \\
370 & 3.03\0(7) & 3.180 & 18.92(7)  & 18.70 & 1.53\0(8)  & 13.2\0(2) & 15.30  \\
385 & 4.12\0(7) & 4.298 & 16.7\0(1) & 16.90 & 2.24\0(9)  & 10.9\0(2) & 13.20  \\
390 & 4.49\0(9) & 4.732 & 15.8\0(2) & 16.20 & 2.5\0\0(1) & 10.2\0(3) & 12.20  \\ \hline
\multicolumn{8}{r}\emph{continued on next page}
\end{tabular}
\end{center}
\end{table}

\begin{table}[ht]
\begin{center}
\begin{tabular}{c|ccccccc}
\multicolumn{8}{l}\emph{continued from previous page} \\
\hline
$T$ & $p_\sigma^{\mathrm{sim}}$
              & $p_\sigma^{\mathrm{exp}}$
                      & $\rho'_{\mathrm{sim}}$
                                  & $\rho'_{\mathrm{exp}}$
                                           & $\rho''_{\mathrm{sim}}$
                                                       & $\Delta h_\mathrm{v}^\mathrm{sim}$
                                                                    & $\Delta h_\mathrm{v}^\mathrm{exp}$ \\
K   & MPa     & MPa   & mol/l      & mol/l   & mol/l     & kJ/mol   & kJ/mol \\ \hline
\multicolumn{8}{l}{Dimethyl Ether} \\ \hline
160 &         &       & 18.25(1)   & 18.209  &         &            &         \\
200 &         &       & 17.17(1)   & 17.189  &         &            &         \\
280 & 0.32(3) & 0.335 & 14.79(1)   & 14.844  & 0.15(1) & 19.90(1)   & 19.896  \\
300 & 0.64(3) & 0.623 & 14.11(1)   & 14.156  & 0.29(1) & 18.57(2)   & 18.623  \\
320 & 1.09(2) & 1.065 & 13.39(1)   & 13.401  & 0.49(1) & 17.11(2)   & 17.124  \\
335 & 1.53(4) & 1.523 & 12.80(2)   & 12.772  & 0.70(1) & 15.86(2)   & 15.806  \\
350 & 2.18(3) & 2.110 & 12.11(3)   & 12.067  & 1.02(1) & 14.29(4)   & 14.262  \\
365 & 2.83(3) & 2.843 & 11.27(3)   & 11.242  & 1.36(2) & 12.65(5)   & 12.391  \\
380 & 3.86(3) & 3.742 & 10.31(7)   & 10.196  & 2.09(3) & 10.1\0(1)  & \09.950 \\
390 & 4.57(3) & 4.444 & \09.6\0(2) & \09.215 & 2.69(4) & \08.3\0(1) & \07.625 \\ \hline
\multicolumn{8}{l}{Sulfur Dioxide} \\ \hline
240 & 0.04(1) & 0.032 & 23.945(3)  & 23.664  & 0.018(4)  & 26.40(1) & 26.197 \\
255 & 0.09(1) & 0.069 & 23.327(3)  & 23.085  & 0.043(4)  & 25.53(1) & 25.375 \\
270 & 0.15(1) & 0.136 & 22.694(3)  & 22.488  & 0.069(4)  & 24.71(1) & 24.521 \\
285 & 0.24(1) & 0.247 & 22.043(4)  & 21.869  & 0.106(6)  & 23.85(1) & 23.623 \\
300 & 0.40(4) & 0.417 & 21.357(8)  & 21.222  & 0.17\0(1) & 22.88(1) & 22.667 \\
330 & 1.03(3) & 1.013 & 19.88\0(1) & 19.816  & 0.44\0(1) & 20.57(1) & 20.511 \\
360 & 2.09(3) & 2.098 & 18.17\0(2) & 18.190  & 0.90\0(2) & 17.83(3) & 17.867 \\
390 & 3.85(4) & 3.873 & 15.98\0(4) & 16.166  & 1.80\0(3) & 14.20(7) & 14.354 \\
405 & 5.17(3) & 5.095 & 14.53\0(6) & 14.855  & 2.66\0(3) & 11.50(7) & 11.967 \\ \hline
\multicolumn{8}{l}{Dimethyl Sulfide} \\ \hline
250 &           &        & 14.385(3)   & 14.390  &           &          &        \\
302 & 0.070(2)  & 0.0747 & 13.448(7)   & 13.492  & 0.028(1)  & 26.43(1) & 27.607 \\
333 & 0.210(5)  & 0.2082 & 12.865(8)   & 12.906  & 0.080(2)  & 25.02(2) & 25.917 \\
363 & 0.47\0(1) & 0.4647 & 12.233(8)   & 12.288  & 0.169(4)  & 23.41(2) & 24.086 \\
383 & 0.74\0(1) & 0.7352 & 11.809(7)   & 11.841  & 0.261(5)  & 22.27(2) & 22.726 \\
403 & 1.11\0(2) & 1.1092 & 11.32\0(1)  & 11.355  & 0.387(7)  & 20.95(2) & 21.216 \\
428 & 1.74\0(2) & 1.7584 & 10.68\0(1)  & 10.673  & 0.609(7)  & 19.07(3) & 19.034 \\
453 & 2.59\0(2) & 2.6616 & \09.95\0(2) & \09.857 & 0.93\0(1) & 16.85(5) & 16.336 \\
478 & 3.68\0(3) & 3.8885 & \09.01\0(2) & \08.770 & 1.39\0(2) & 14.01(8) & 12.580 \\ \hline
\multicolumn{8}{l}{Thiophene} \\ \hline
270 & 0.0009(1)   & 0.0023 & 12.83\0(2)  & 12.967  & 0.0004(1)   & 34.13(7)  & 35.956 \\
285 & 0.0033(2)   & 0.0054 & 12.62\0(2)  & 12.763  & 0.0014(1)   & 33.46(6)  & 35.270 \\
300 & 0.0101(8)   & 0.0115 & 12.398(6)   & 12.555  & 0.0041(3)   & 32.72(2)  & 34.562 \\
325 & 0.029\0(2)  & 0.0333 & 12.068(6)   & 12.199  & 0.0110(9)   & 31.67(2)  & 33.329 \\
348 & 0.076\0(5)  & 0.0753 & 11.732(6)   & 11.859  & 0.027\0(2)  & 30.59(2)  & 32.127 \\
383 & 0.19\0\0(3) & 0.2089 & 11.257(8)   & 11.315  & 0.063\0(9)  & 29.06(2)  & 30.148 \\
418 & 0.53\0\0(2) & 0.4755 & 10.699(6)   & 10.728  & 0.166\0(5)  & 27.02(2)  & 27.939 \\
464 & 1.18\0\0(2) & 1.1389 & \09.88\0(1) & \09.866 & 0.359\0(6)  & 24.09(3)  & 24.532 \\
493 & 1.94\0\0(2) & 1.8132 & \09.33\0(2) & \09.238 & 0.599\0(8)  & 21.76(5)  & 21.928 \\
530 & 3.17\0\0(2) & 3.0679 & \08.42\0(3) & \08.264 & 1.03\0\0(1) & 18.06(7)  & 17.653 \\
551 & 4.20\0\0(3) & 4.0312 & \07.86\0(5) & \07.530 & 1.46\0\0(2) & 15.2\0(1) & 14.240 \\ \hline
\multicolumn{8}{r}\emph{continued on next page}
\end{tabular}
\end{center}
\end{table}

\begin{table}[ht]
\begin{center}
\begin{tabular}{c|ccccccc}
\multicolumn{8}{l}\emph{continued from previous page} \\
\hline
$T$ & $p_\sigma^{\mathrm{sim}}$
              & $p_\sigma^{\mathrm{exp}}$
                      & $\rho'_{\mathrm{sim}}$
                                  & $\rho'_{\mathrm{exp}}$
                                           & $\rho''_{\mathrm{sim}}$
                                                       & $\Delta h_\mathrm{v}^\mathrm{sim}$
                                                                    & $\Delta h_\mathrm{v}^\mathrm{exp}$ \\
K   & MPa       & MPa    & mol/l         & mol/l   & mol/l     & kJ/mol   & kJ/mol \\ \hline
\multicolumn{8}{l}{Hydrogen Cyanide} \\ \hline
273 & 0.030(1)  & 0.0350 & 26.21(3)  & 26.548 & 0.0141(4)   & 28.78(4)  &  27.778  \\
295 & 0.078(3)  & 0.0877 & 25.11(4)  & 25.354 & 0.035\0(1)  & 27.30(4)  &  27.060  \\
315 & 0.169(5)  & 0.1798 & 24.01(5)  & 24.217 & 0.074\0(2)  & 25.76(5)  &  26.336  \\
358 & 0.66\0(1) & 0.6400 & 21.52(8)  & 21.536 & 0.292\0(6)  & 21.86(7)  &  24.451  \\
388 & 1.43\0(3) & 1.3255 & 19.52(7)  & 19.367 & 0.69\0\0(1) & 18.41(8)  &  22.697  \\
406 & 2.09\0(4) & 1.9655 & 17.9\0(8) & 17.859 & 1.08\0\0(2) & 16.0\0(1) &  21.323  \\
420 & 2.69\0(3) & 2.6261 & 16.3\0(1) & 16.504 & 1.44\0\0(3) & 14.1\0(1) &  19.953  \\
430 & 3.43\0(3) & 3.2072 & 15.5\0(1) & 15.373 & 2.05\0\0(3) & 12.1\0(1) &  18.690  \\
435 & 3.70\0(3) & 3.5377 & 14.0\0(3) & 14.726 & 2.31\0\0(3) & 10.7\0(2) &  17.909  \\ \hline
\multicolumn{8}{l}{Acetonitrile} \\ \hline
270 & 0.0015(1)   & 0.0029  & 19.71(1) & 19.636 & 0.0007(1)   & 38.87(8)  & 34.302 \\
300 & 0.0074(3)   & 0.0132  & 18.90(2) & 18.871 & 0.0032(1)   & 36.78(7)  & 32.952 \\
327 & 0.029\0(1)  & 0.0396  & 18.20(3) & 18.151 & 0.0112(6)   & 34.88(7)  & 31.641 \\
360 & 0.088\0(3)  & 0.119\0 & 17.25(2) & 17.223 & 0.033\0(1)  & 32.42(6)  & 29.886 \\
400 & 0.294\0(5)  & 0.352\0 & 16.05(2) & 16.003 & 0.109\0(2)  & 29.09(6)  & 27.459 \\
420 & 0.51\0\0(1) & 0.558\0 & 15.47(4) & 15.341 & 0.190\0(4)  & 27.27(8)  & 26.078 \\
436 & 0.76\0\0(3) & 0.782\0 & 14.78(1) & 14.777 & 0.289\0(9)  & 25.31(5)  & 24.866 \\
450 & 1.00\0\0(4) & 1.032\0 & 14.25(2) & 14.253 & 0.38\0\0(1) & 23.98(5)  & 23.705 \\
464 & 1.27\0\0(9) & 1.341\0 & 13.65(3) & 13.693 & 0.48\0\0(3) & 22.58(5)  & 22.427 \\
490 & 2.08\0\0(4) & 2.101\0 & 12.42(3) & 12.515 & 0.85\0\0(2) & 19.20(9)  & 19.603 \\
505 & 2.69\0\0(3) & 2.673\0 & 11.52(5) & 11.707 & 1.18\0\0(2) & 16.7\0(1) & 17.549 \\
510 & 2.73\0\0(5) & 2.889\0 & 11.06(6) & 11.405 & 1.16\0\0(2) & 16.5\0(1) & 16.753 \\
518 & 3.14\0\0(4) & 3.263\0 & 10.37(7) & 10.871 & 1.37\0\0(2) & 15.0\0(1) & 15.308 \\ \hline
\multicolumn{8}{l}{Nitromethane} \\ \hline
290 & 0.001(1)  & 0.0031 & 18.82(2)  & 18.669 & 0.0003(1)   & 44.05(4)  & 38.511 \\
353 & 0.026(1)  & 0.0503 & 17.25(2)  & 17.239 & 0.0092(4)   & 39.62(5)  & 35.812 \\
390 & 0.105(3)  & 0.1602 & 16.32(3)  & 16.331 & 0.035\0(1)  & 36.78(7)  & 33.983 \\
430 & 0.33\0(1) & 0.4354 & 15.27(2)  & 15.270 & 0.107\0(3)  & 33.47(5)  & 31.716 \\
471 & 0.92\0(2) & 1.0033 & 14.13(3)  & 14.058 & 0.301\0(6)  & 29.29(8)  & 28.931 \\
500 & 1.56\0(2) & 1.6696 & 13.12(4)  & 13.083 & 0.525\0(8)  & 25.98(9)  & 26.518 \\
530 & 2.70\0(4) & 2.6942 & 11.93(5)  & 11.899 & 0.99\0\0(2) & 21.5\0(1) & 23.347 \\
560 & 4.09\0(6) & 4.2110 & 10.4\0(1) & 10.345 & 1.64\0\0(3) & 16.7\0(2) & 18.696 \\ \hline\hline
\end{tabular}
\end{center}
\end{table}

\end{document}